\documentclass[preprint,12pt]{elsarticle}

\usepackage{amsmath, amssymb}

\usepackage{xcolor}
\usepackage[numbers]{natbib}

\usepackage{amscd}
\usepackage{amsfonts}
\usepackage{fancyhdr}
\usepackage{hyperref}
\usepackage{subfig}
\usepackage{float}
\usepackage{multimedia}
\usepackage{microtype}
\usepackage{tikz}
\usepackage{graphicx}      % include this line if your document contains figures
\graphicspath{{1stJournalImage/}}

\usepackage{tikz}
\usetikzlibrary{shapes.geometric, arrows}
\usetikzlibrary{arrows,calc,positioning}
\usepackage{float}
\usepackage{algorithm2e}
\usepackage{multimedia}
\usepackage[linguistics]{forest} 
\usepackage{lmodern}
\usepackage{microtype}

\tikzstyle{intt}=[draw,text centered,minimum size=6em,text width=5.25cm,text height=0.34cm]
\tikzstyle{intl}=[draw,text centered,minimum size=2em,text width=2.75cm,text height=0.34cm]
\tikzstyle{int}=[draw,minimum size=2.5em,text centered,text width=3.5cm]
\tikzstyle{intg}=[draw,minimum size=3em,text centered,text width=6.cm]
\tikzstyle{sum}=[draw,shape=circle,inner sep=2pt,text centered,node distance=3.5cm]
\tikzstyle{summ}=[drawshape=circle,inner sep=4pt,text centered,node distance=3.cm]
\tikzstyle{empty} = [draw]

\usepackage{amssymb}
\usepackage{amsmath}
\journal{Nuclear Physics B}

\begin{document}

\begin{frontmatter}



\title{Calibration of a Macroscopic Coupled People-Epidemic Transport PDE Model via Density and Velocity Computation from Microscopic Data}


\author{Nikoletta Hadjihabi}  %% Author name
\author{Nikolaos Bekiaris-Liberis}  

\affiliation{organization={Department of Electrical and Computer Engineering, Technical 
		University of Crete},%Department and Organization
	addressline={University Campus}, 
	city={Chania},
	postcode={73100},
	country={Greece}}

%% Abstract
\begin{abstract}
We introduce an approach for derivation and smoothing of macroscopic densities and velocities from people trajectories, obtained from microscopic position data of individuals within a football stadium. We compute macroscopic densities specifically for susceptible, infected, and exposed individuals, via detection of exposed individuals based on the duration of critical contacts between susceptible and infected. We then present and numerically solve a crowd flow - epidemic spreading PDE model using a finite volume scheme. Finally, we calibrate the model using the smoothed densities computed.
\end{abstract}



%% Keywords

\end{frontmatter}

\section{Introduction}
\subsection{Motivation}
In many real-world scenarios, human mobility and epidemic spread are interconnected dynamic processes. The movement of individuals influences the transmission of infectious diseases, while outbreaks can alter mobility patterns \cite{transmissionOutbreaks}. Modeling these coupled systems is critical for applications such as epidemic control and public health policy design \cite{OptimalControl}. Macroscopic crowd flow -- epidemic spreading modeling is essential when examining large-scale movement patterns such as, for example, within a stadium \cite{LargeScaleEvents}, as it provides a computationally tractable approach for analyzing the dynamics of  key quantities, such as density and speed of individuals \cite{HelbingCrowdDisaster}. For a more realistic description of such macroscopic dynamics, a model employed should be calibrated using real microscopic data \cite{AnalysisEmpiricalTrajectory}. Effectively calibrating a macroscopic coupled crowd--epidemic model requires proper processing of available data, as raw data capturing crowd flow and epidemic spreading dynamics in a given space are either unavailable or very sparse, at least on a time scale of minutes and a spatial scale of meters \cite{dataGaps}. Consequently, the above motivate our study on computation of macroscopic quantities from available people trajectories data (taken from a stadium over a duration of minutes) and their utilization in calibration of a macroscopic, coupled people-epidemic transport model.

%Additionally, controlled experiments with volunteers provide another approach, offering valuable insights. However, these experiments typically involve smaller sample sizes and may not fully capture the diversity of behaviors seen in real-world pedestrian environments. 
%%%%%%%%%%%%%%%%%%%%%%%\cite{inbook}%%%%%%%%%%%%%%%%%%%%%%%
\subsection{Literature on Macroscopic People-Epidemic Transport Models and Their Calibration}
Macroscopic quantities may provide holistic insights into the overall influence of human mobility on epidemic spreading, when modeling epidemic transport through individual--to--individual interactions may becomes intractable, such as, for example, in the case of large-scale events, e.g., a football match or a concert \cite{LargeScaleEvents}. On the other hand, reliable computation of macroscopic quantities, such as density and velocity of people in susceptible, infected, or exposed categories, based on individuals' trajectories is challenging. This challenge arises either when utilizing a macroscopic model calibrated with microscopic data or when directly processing available data, due to the general unavailability of both people's trajectories and corresponding epidemiological data within the same space--time domain \cite{dataGaps}. The sparsity of the data requires development and utilization of appropriate techniques to compute smooth macroscopic densities and velocities such as, for example, kernel-based methods, as it has been successfully done in traffic flow-oriented studies, see, e.g., \cite{REMPE2017644, TUMASH2022374}. 

The study of crowd flow has progressed considerably over time \cite{ BellomoDogbeChristian, PedestrianDynamics, Goatin2009, XiaEtEl, Cristiani2014, ComparativeMacroscopicPed, MAITY2024205}. While pedestrian motion is often linked to vehicle traffic flow models, which are typically treated as one-dimensional problems, crowd flow inherently involves two-dimensional dynamics, as discussed in \cite{inproceedings, Hugens}. However, some studies, such as the ones in \cite{GOTTLICH201536, Herty2018, mollierPaper, tumash:tel-03474112, KARAFYLLIS2022110517, Treiber2024}, extend traffic flow theory to multi-directional scenarios on $2$-D domains. Moreover, the connection between crowd movement and epidemic spreading in certain spaces, is explored in \cite{Bertaglia2020, DiseaseContagion, AbdulSalam2023, Angelli2023, delis}, using partial differential equation (PDE)-based macroscopic models of coupled crowd flow -- epidemics spreading processes. On a higher, macroscopic level, e.g., on a city or country level, coupled human mobility -- disease transmission models using Ordinary Differential Equations (ODEs) are presented, for example, in \cite{Arino2004, 6113095, OptimalControl}.  Despite availability of macroscopic PDE models of crowd flow and crowd-epidemic transport dynamics, to the best of our knowledge, there exists no result addressing their calibration with real microscopic data. The available relevant results concern calibration of a first-order, PDE crowd flow model, using fictitious data generated by a microscopic simulator \cite{Parameter_estimation}; or calibration of microscopic ODE models, using real microscopic data \cite{Calibration_Bayesian_Optimization, China_2013, Bode_2020}.%The only available relevant results are those concerning calibration of ODE models for pedestrians movement using data generated by microscopic simulators. 

\subsection{Contributions}
The first key contribution of this paper lies in derivation and smoothing of macroscopic densities and velocities from individual people trajectories from the data in \cite{LargeScaleEvents}, corresponding to a football match, discerning between density of exposed, infected, or susceptible individuals. First we compute the trajectories of exposed individuals from the trajectories of all people, by analyzing the movements of infected individuals (randomly assigned from the crowd) and identifying their critical contacts. Specifically, we classify a susceptible individual as exposed based on computation of an infection probability, which depends on the cumulative contact duration between infected and susceptible individuals. We then compute the macroscopic densities of the total population, susceptible individuals, and exposed individuals at each time step by using the number of individuals in a group located within each square area into which the spatial domain is discretized. Due to the sparsity of data and to assign a density value throughout the computational domain (making the densities computed useful for model calibration purposes) we also apply smoothing. To obtain smoothed macroscopic densities, we calculate, for each group, a two-dimensional probability density function (PDF) using Gaussian kernels. 

Next, we compute the macroscopic speed magnitudes for each cell, as the square root of the sum of the squares of the average speeds in the $x-$ and $y-$directions, which in turn are computed averaging (in each $x-$ and $y-$direction) the speed magnitude of each individual in the crowd. The speed direction is consequently also derived from the angle formed between the speed components in each direction. We then compute the smoothed macroscopic speed components in each direction employing Gaussian kernels as weights on the data corresponding to each individual's speed components. Finally, the direction of smoothed speed is determined from the angle formed by the smoothed speed components in the $x-$ and $y-$directions.

The second key contribution lies in formulating and numerically solving a coupled people-epidemic PDE model, and in its calibration using the computed macroscopic densities. The crowd flow model consists of a continuity equation for mass conservation coupled with momentum equations in each spatial direction. The crowd flow model is coupled with an epidemic PDE model describing density dynamics of susceptible, exposed, and infected individuals. Disease transmision is modeled through a source term that transfers individuals from susceptible to the exposed group at a specific infection rate, whose computation is dictated by the percentage of infected individuals, the average distances betweeen pedestrians, and their average relative speeds. For the numerical solution of the model, we employ a standard finite volume scheme. Next, we perform calibration of the model, starting with calibration of the crowd flow component as the key step. We use the Nelder-Mead algorithm with a cost function the root mean squared error (RMSE) between predicted and smoothed observed densities in the spatial domain considered. The calibration procedure results, on average, in about $24\%$ relative error between model-predicted and actual densities, which is deemed adequate given the sparsity of available data. We then perform calibration of the complete model using densities of exposed individuals as well.

The present journal paper extends our preliminary conference version \cite{HADJIHABI}, in that here we a) present and solve numerically a coupled crowd flow - epidemic spreading PDE model, and  b) utilize the macroscopic data computed in \cite{HADJIHABI} for its calibration.

\subsection{Organization}
The structure of the paper is as follows. In Section \ref{ComputationOfPeopleTrajectoriesFromTheData}, we present the process for computing exposed individuals' trajectories from the data. In Section \ref{ComputationOfMacroscopicDensitiesAndSpeeds}, we introduce computation of macroscopic densities and velocities, as well as their smoothed counterparts. In Section \ref{Model_presentation}, we present the coupled people-epidemic transport model and, in Section \ref{Numerical_scheme_presentation}, we introduce the numerical scheme used to solve it. In Section \ref{Model_calibration}, we introduce the calibration procedure and present the respective results. In Section \ref{Remarks} we provide concluding remarks.
\section{Computation of Exposed People Trajectories From Data} \label{ComputationOfPeopleTrajectoriesFromTheData}
\subsection{Data Description} \label{Data_description}
\begin{figure} [ht]
	\centering
	\includegraphics[width=0.7\linewidth]{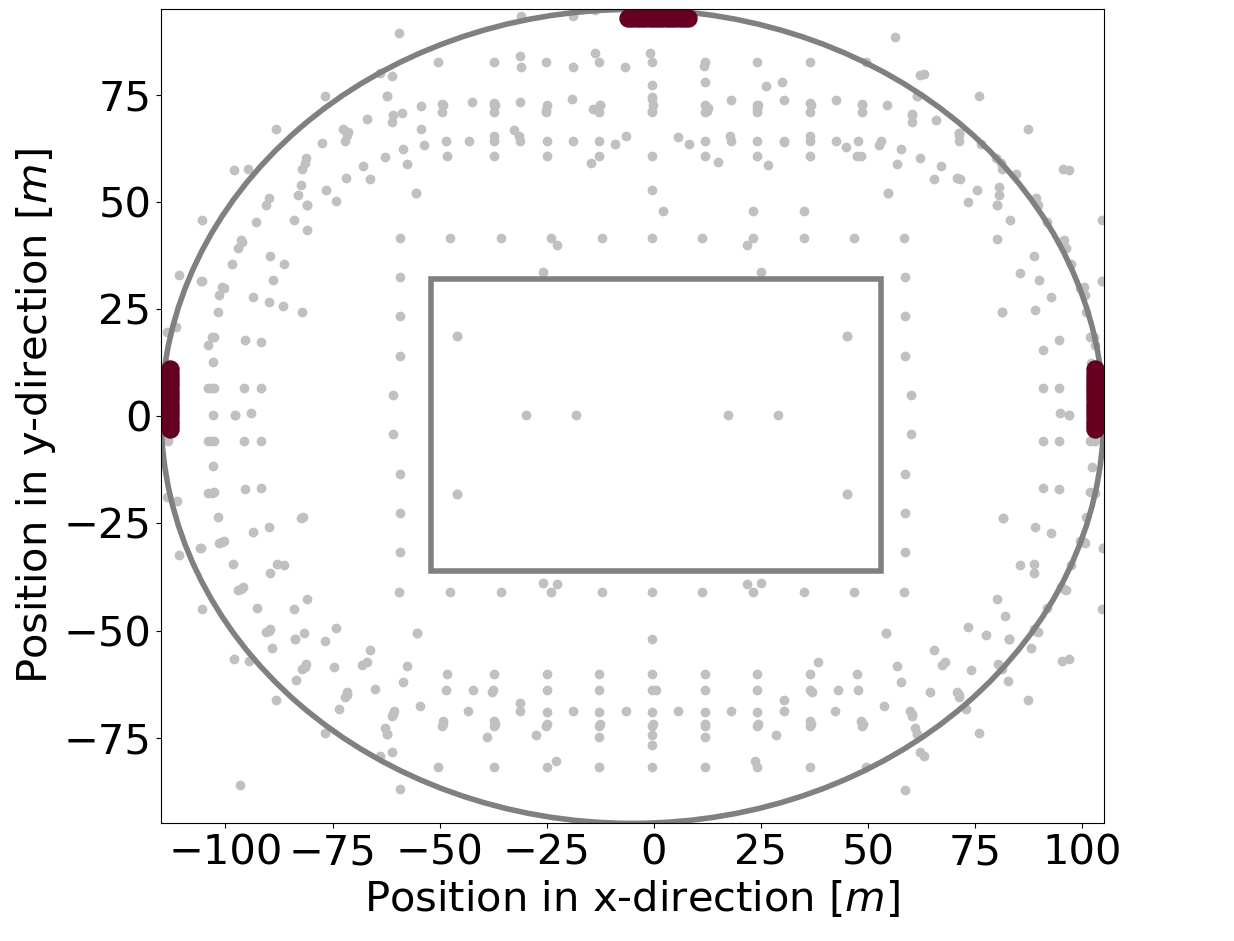} 
	\caption{Access points throughout the observed region $\Omega = [-115,105] \times [-95,95]$. The exits are also shown.}
	\label{Points_entrances}
\end{figure}

The following results are derived using the data that are collected by Wi-Fi network in Johan Cruijff Arena during the football match Ajax--Feyenoord in October 2019, as they are given in \cite{LargeScaleEvents}. The stadium has dimensions $226  \text{ m}  \times 190 \text{ m}  \times 72 \text{ m}$ while the field has dimensions  $105  \text{ m} \times 68  \text{ m}. $ The stadium's capacity is $56120$ visitors during a football match. The entrances/exits are located on the north, east, and west sides of the stadium as they are shown in Figure \ref{Points_entrances}. In addition, the layouts of both the stadium's and field's borders are shown in the same figure. The observed region for our computations is $\Omega = [-115, 105] \times [-95,95]$. 

During the football match, the tracking time started at 16:00 and finished at 20:00. The match started at 16:45 and it consisted of two 45-minute halves with a 15-minute break in between. The data contain only the positions of 362 individuals out of $53968$ at different time instances. The time instances range from $0$ to $1440$ and each time instant correspond to $10$ seconds.  

The spatial coordinates of individuals' position are derived from Wi-Fi detections. The wireless network in the stadium consists of 591 access points with known spatial coordinates. 
Figure \ref{Points_entrances} shows the distribution of access points, where each point in this figure represents a distinct access point. The position of each individual at a specific time instance is computed using proximity detection. More details about the data collection are described in \cite{ref25}.
%Johan Cruijff Arena has a capacity of $56,120$ visitors during a football match and it has a capacity of $71,000$ visitors during a music concert. 

\subsection{Critical Contacts Description} 
We focus on a basic susceptible--exposed model and do not include the transitions from exposed to infected, or from infected to recovered, due to the time scales considered. We divide the population in three states: Susceptible (S), Exposed (E), and Infected (I). The infected agents are selected randomly. The growth in the number of exposed individuals is influenced by two factors. The first is that the distance between a susceptible individual and an infected one must be less than $1.5$ meters, which is defined as a critical contact. The second factor depends on the definition of the probability of infection $p_{ij}$, which is discussed in Section \ref{Computation_of_trajectories} and depends on the duration of a critical contact. While the total number of exposed individuals increases, the number of infected individuals remains unaffected due to the small time-scales considered.   

We assume that infected individuals enter the stadium at $\bar{t}=9000 \text{ s}$, which we will use as the reference starting point. We select $\bar{t}=9000 \text{ s}$ corresponding to $18:30$, because at this moment there is a continuous movement towards the exits, resulting in more frequent and prolonged contacts, as the match ended at $18:30$. From this reference time until $20:00$, we calculate the critical contacts between susceptible and infected individuals. Thus, we can compute the trajectories of exposed individuals during this interval, as well as the trajectories of susceptible and infected individuals. For the results presented, we will shift the timeline so that $\bar{t} = 9000 \text{ s}$ is represented as $t=0 \text{ s}$, simplifying the timeline for analysis and results' exposition.

\subsection{Computation of Trajectories of Exposed Individuals} \label{Computation_of_trajectories}
We consider that the probability of infection depends on the duration of contact time, i.e., the longer the contact between infected and susceptible individuals, the greater the likelihood of exposure. %Conversely, prolonged contact between susceptible individuals increases the chance of avoiding infection. 
Based on the trajectories of infected individuals, we identify their critical contacts and determine the total duration of each contact. We introduce an infection probability, as described in \cite{LargeScaleEvents}, as a function of contact duration, defined as 
\begin{equation} \label{p_infection}
	p_{ij} = 1 - e^{-\gamma t_{ij}},
\end{equation}
where $t_{ij}$ is the cumulative contact duration between infected individual $i$ and susceptible individual $j$, and $\gamma$ is a parameter modulating the increase of $p_{ij}$ with $t_{ij}$. The variable $e^{-\gamma t_{ij}}$ represents the probability that an individual does not become exposed. Furthermore, as introduced in \cite{modelRiley}, when examining infection transmission within a room, parameter $\gamma$ can be represented as
\begin{equation} \label{gamma_def}
	\gamma = \frac{I_0qp}{Q},
\end{equation}
where $I_0, p, q,$ and $Q$ are defined as follows. The parameter $I_0$ represents the number of initially infected individuals, $p$ denotes the total volume of air inhaled or exhaled by an individual, $q$ is the amount of infections dose released into the air by an infected person, and $Q$ represents the room ventilation rate. In the case of Johan Cruijff Arena during the specific football match, ventilation occurred naturally through the open roof and small openings in the stadium envelope. Due to this, $Q$ represents the volumetric air flow rate into the enclosure, which can be calculated by multiplying the total area of the openings, through which air flows, with the velocity of the air entering through these openings, see \cite{StudyForAmsterdamArena}. The values of $p$ and $q$ in $(\ref{gamma_def})$ were chosen based on the parameters used in \cite{modelRiley} and \cite{StudyForAmsterdamArena}, and they are set to $p=9.433 \times 10^{-4} \text{ m}^3 \text{s}^{-1}$, $q=3 \times 10^{-3} \text{ s}^{-1} \text{ per infected person}$, $Q=4.875 \times 10^{-2} \text{ m}^3 \text{s}^{-1}$, and $I_0 = 54 \text{ infected people}$.  

\begin{figure} [H]
	\centering
	\includegraphics[width=0.6\linewidth]{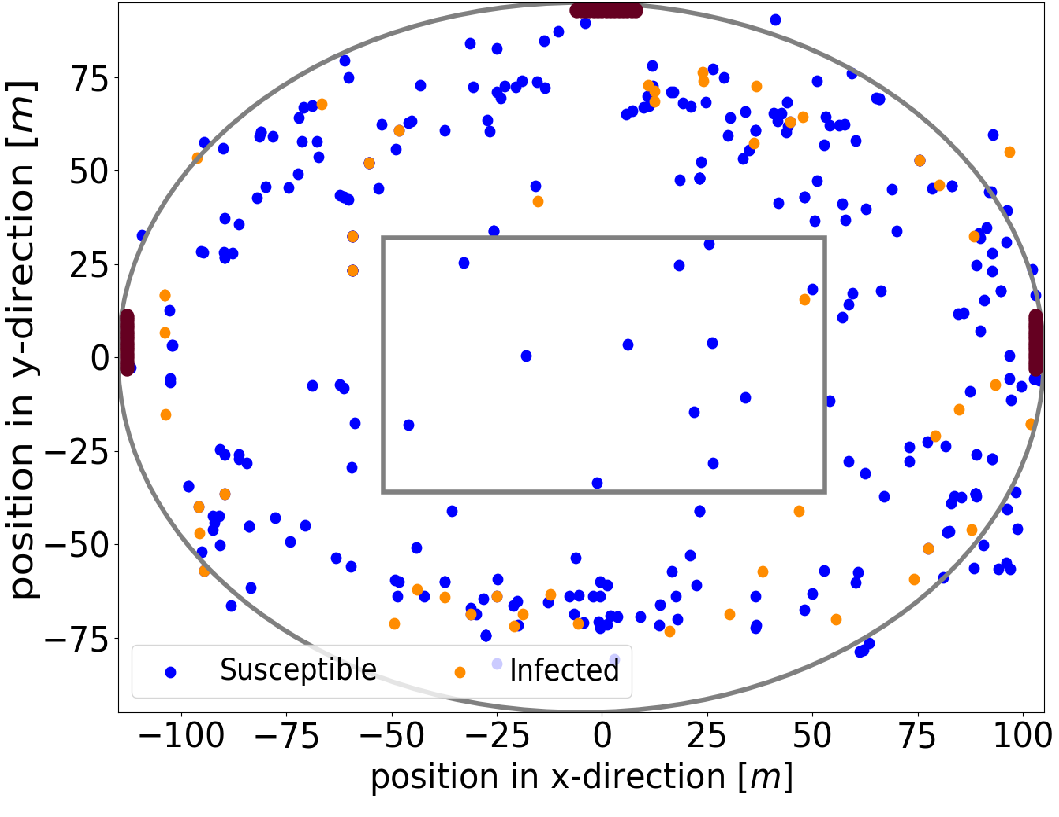}
	\caption{Position of individuals from the data at the initial time when there are no exposed individuals, $298$ susceptible and $52$ infected individuals.}
	\label{exposed_0}
\end{figure}

\begin{figure} [H]
	\centering
	\includegraphics[width=0.6\linewidth]{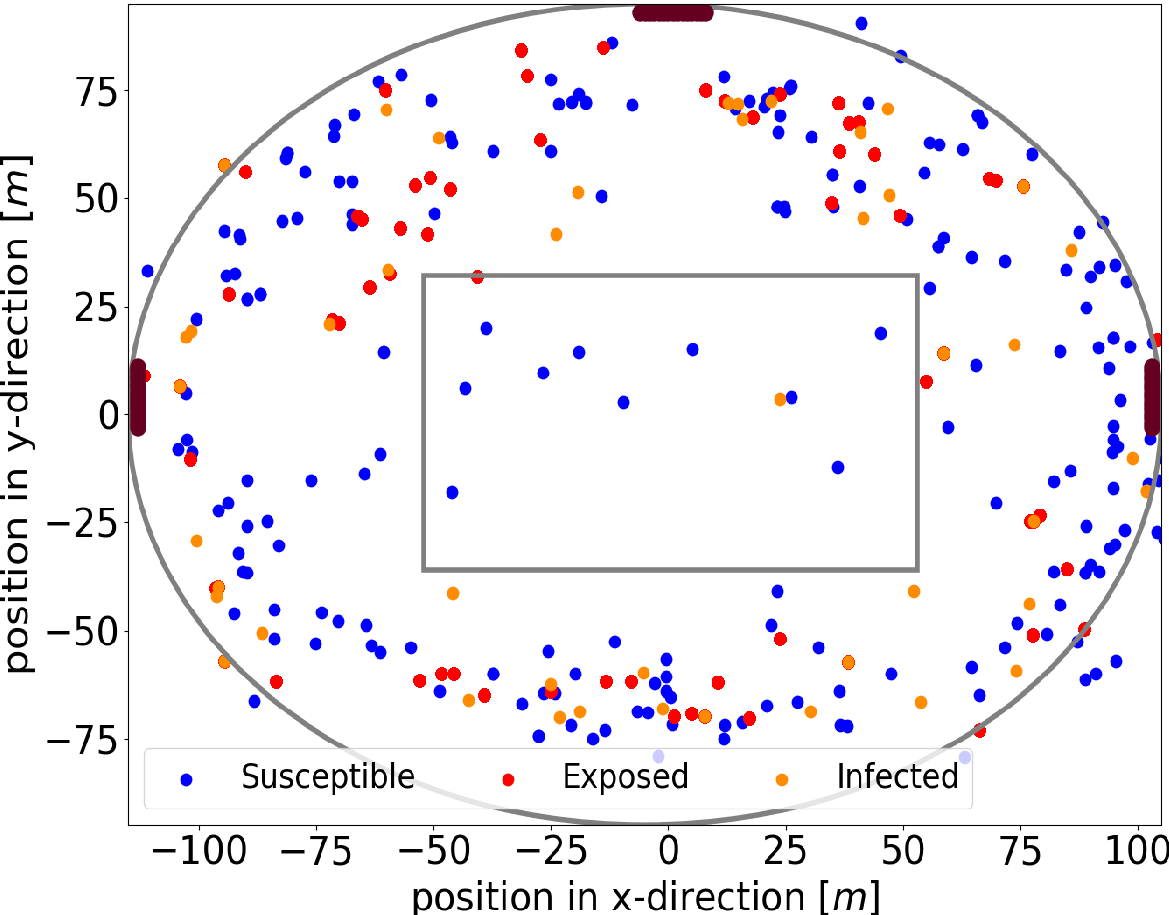}
	\caption{Position of individuals from the data at $t=450 \text{ s}$ when there are 38 exposed, 243 susceptible individuals, and 46 infected individuals.}
	\label{exposed_pos_945}
\end{figure}

To identify which susceptible individuals become exposed, we utilize a process based on the Bernoulli distribution. The process is described as follows.
At each time step, for every pair of susceptible (denoted by $j$) and infected (denoted by $i$) individuals, a random number is generated  within the interval $(0,1)$. This random number is compared to the probability $1-p_{ij}$, where $p_{ij}$ is calculated by (\ref{p_infection}). The probability $1-p_{ij}$ is used as a threshold to determine whether the susceptible individual becomes exposed. If the generated random number is greater than or equal to $1-p_{ij}$, the susceptible individual is considered exposed. Using this method to track exposed individuals, Figures \ref{exposed_0} and \ref{exposed_pos_945} show the positions of infected, exposed, and susceptible individuals at two time instances. Figure \ref{exposed_pos_945} indicates an increase in the number of exposed individuals as infected individuals move within the crowd, while the locations of exposure depend on the location of infected individuals. Figure \ref{exposed_pos_945} shows a general decrease in the population within the stadium, which is attributed to the fact that individuals exit the stadium after the match has ended. 

In Figures \ref{exposed_0} and \ref{exposed_pos_945}, we also observe individuals detected in the field, who are probably security personnel, maintenance crew, technicians, or similar staff. Additionally, we observe individuals detected outside the stadium, as access points are also located in areas surrounding the stadium.

\section{Computation of Macroscopic Densities and Velocities} \label{ComputationOfMacroscopicDensitiesAndSpeeds}
\subsection{Macroscopic Densities Computation} \label{Macroscopic_densities}
%%%% what I had in section 6 %%%%
The stadium is divided into squares of an area of $A=5 \text{ m}^2$ each, for improved presentation clarity in the plots. We note that due to the sparsity of available data, the results obtained using squares with an area of $A=5 \text{ m}^2$ are quite similar to those obtained using squares of any area $A$ such that $1 \text{ m}^2 \leq A \leq 5 \text{ m}^2$. For each time instant and based on the data described in Section \ref{Data_description}, we compute the number of individuals at each square. The density at time $t$ is given by
\begin{equation} \label{Density_estimation}
	\bar{\rho}(x,y,t)= \frac{n(x,y,t)}{A},
\end{equation}
where $n(x,y,t)$ is the total number of individuals within the square centered at $(x,y)$ at time $t$. We compute the density of each group (susceptible and exposed individuals, as the density of infected individuals is then derived from the total density) similarly to equation (\ref{Density_estimation}). The modified formulas for calculating the density of exposed and susceptible individuals are 
\begin{equation} \label{Density_exposed_susceptible_estimation}
	\bar{\rho}^\text{E} (x,y,t) = \frac{n^\text{E}(x,y,t)}{A}, \hspace{3mm} \bar{\rho}^\text{S} (x,y,t) = \frac{n^\text{S}(x,y,t)}{A},
\end{equation}
where $n^\text{E}(x,y,t)$ and $n^\text{S}(x,y,t)$ represent the total number of exposed and susceptible individuals, respectively, at time $t$.

\begin{figure} [H]
	\centering
	\includegraphics[width=0.65\linewidth]{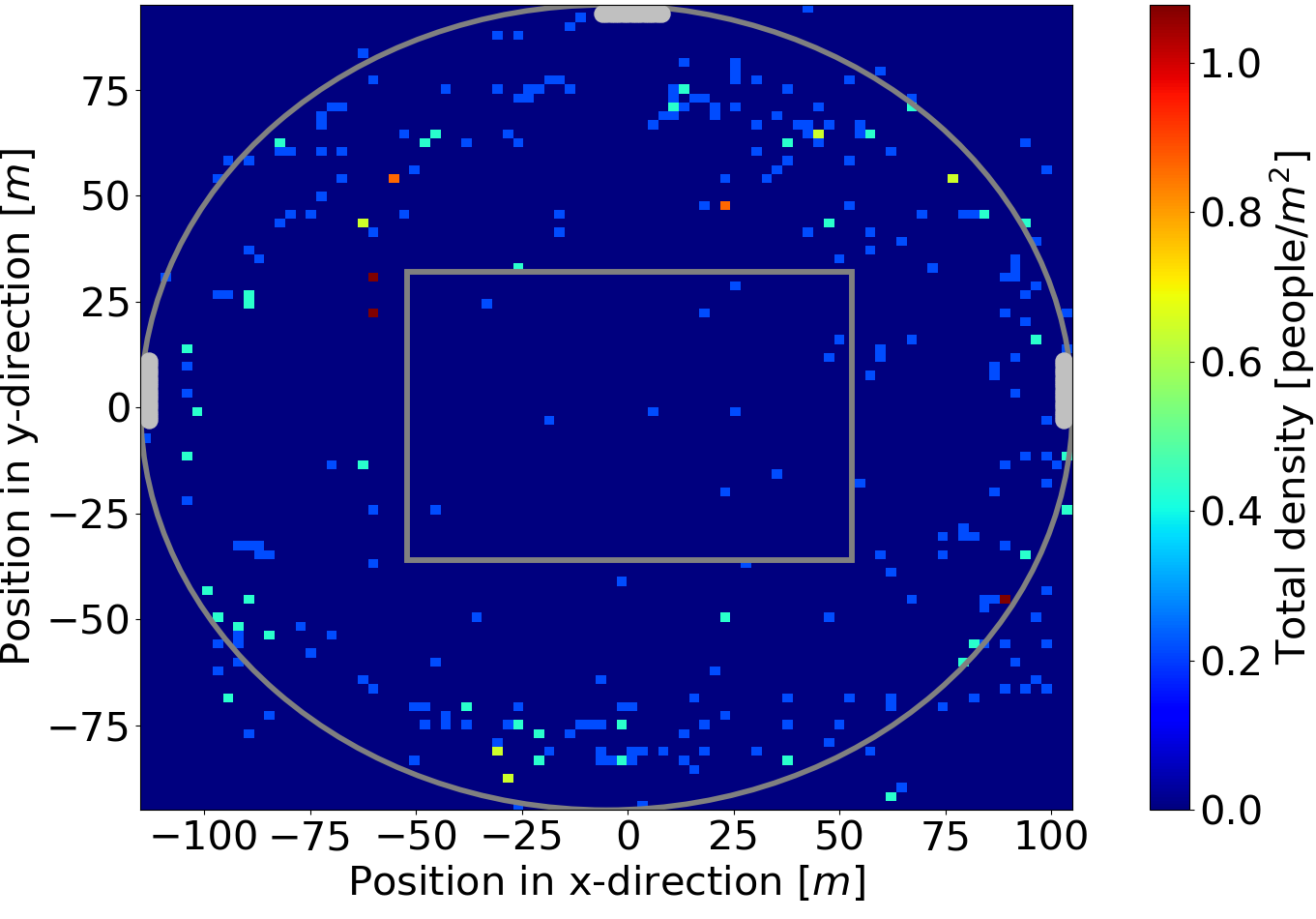}
	\caption{Total density at the initial time computed from the data using equation (\ref{Density_estimation}).} 
	\label{total_density_hist_901}
\end{figure}

\begin{figure} [H]
	\centering
	\includegraphics[width=0.65\linewidth]{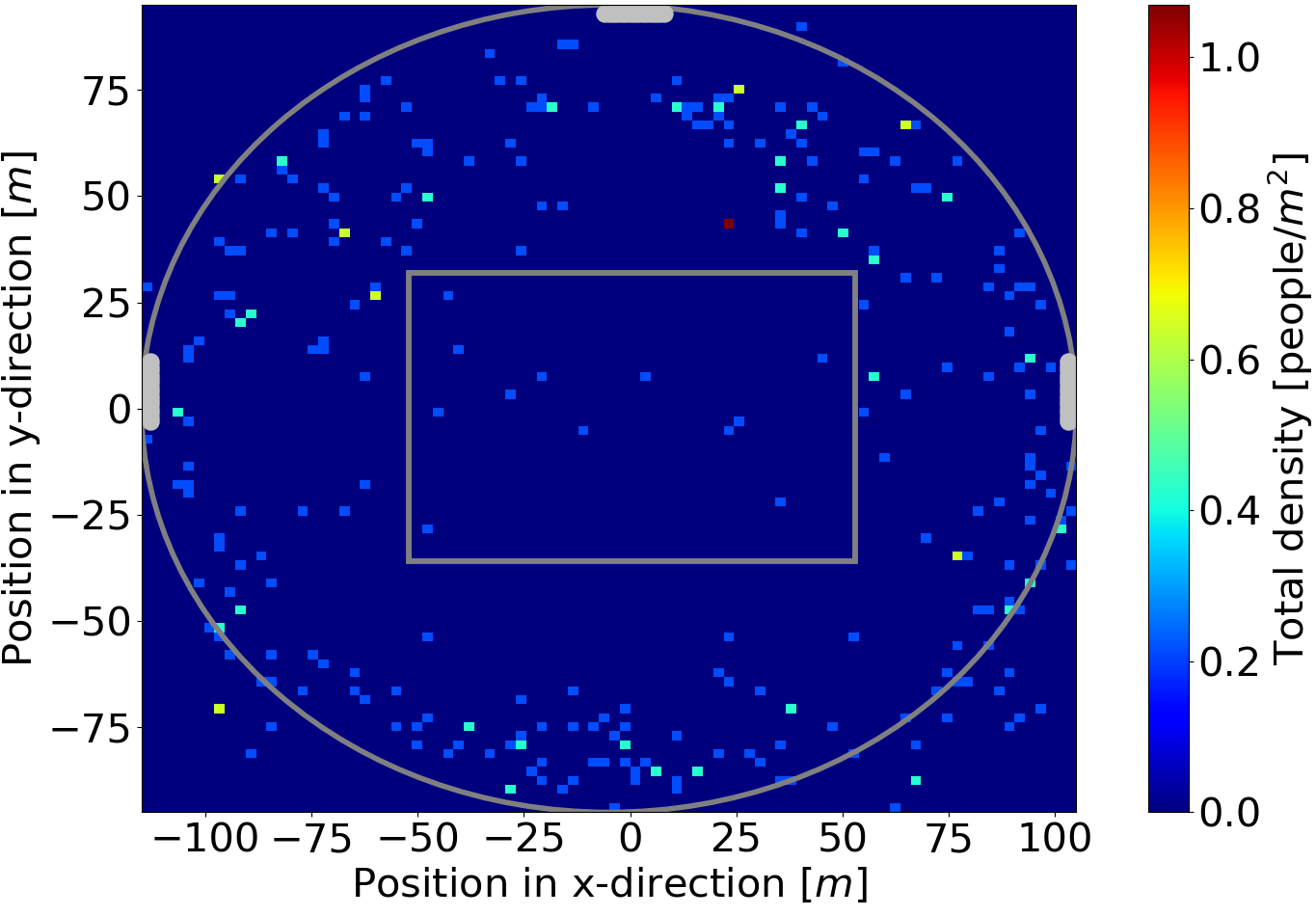}
	\caption{Total density at $t=450 \text{ s}$ computed from the data using equation (\ref{Density_estimation}).}
	\label{total_density_hist_945}
\end{figure}

\begin{figure}[H]
	\centering
	\includegraphics[width=0.65\linewidth]{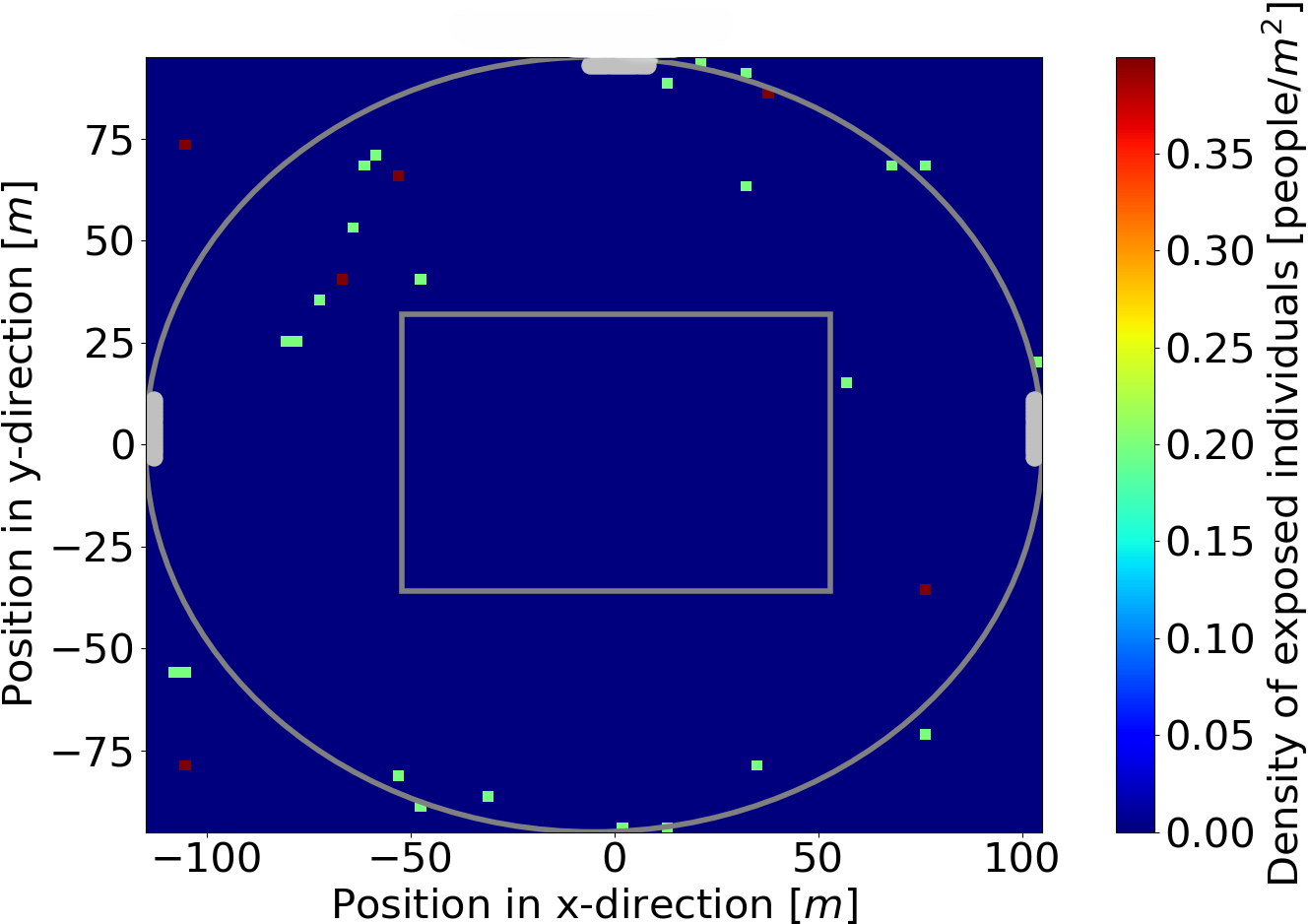}
	\caption{Density of exposed individuals at $t=450 \text{ s}$ computed from the data using equation (\ref{Density_exposed_susceptible_estimation}).}
	\label{exposed_density_hist_945}
\end{figure}
Figures \ref{total_density_hist_901} and \ref{total_density_hist_945} show the total density at the time instances $t =0 \text{ s}$ and $t=450 \text{ s}$, while Figure \ref{exposed_density_hist_945} shows the density of exposed individuals at $t=450 \text{ s}$. We observe in particular a nonzero density in the region of stadium's field. This is likely due to the presence of security personnel or other stadium staff after the match, who are detected inside the field. We observe that due to scarcity of available data, density values are not assigned throughout the computational domain, which calls for smoothing.

For a smoother density computation and to assign density values throughout the computational domain, we use the Kernel Density Estimation (KDE) method, see, e.g., \cite{mollier:tel-02905489}. This ensures the computed quantities are useful for calibrating a macroscopic model as the one in \cite{DiseaseContagion}.  KDE is a non-parametric technique used to estimate the Probability Density Function (PDF) of a random variable. The process involves selecting a kernel function to smooth the data points across the observed region. In the present case, we choose the Gaussian kernel function defined as 
\begin{equation}\label{Gaussian_kernel}
	K(u) = \frac{1}{\sqrt{2 \pi}}{e^{-\frac{u^2}{2}}}.
\end{equation}
Each data point contributes to the estimated density based on its proximity to other points, with the influence of each point decreasing with distance. This decrease is controlled by bandwidth parameters $h_x$ and $h_y$, which define the width of the neighborhood around each point in the $x-$ and $y-$directions, respectively, see also, e.g., \cite{ElementsStatistical}. Using this approach, we compute the two-dimensional PDF defined as
\begin{equation} \label{PDF_function_2D}
	\hat{f}(x, y, t) = \frac{1}{N(t)} \sum_{i=1}^{N(t)} K_{h_x}\left(x_{i,t} - x\right) K_{h_y}\left(y_{i,t} - y\right),
\end{equation}
where
\begin{equation} 
	\left.\begin{aligned}
		K_{h_x}\left(x_{i,t} - x\right) & = \frac{1}{h_x} K\left( \frac{x_{i,t} - x }{h_x} \right), \\
		K_{h_y}\left(y_{i,t} - y\right) & = \frac{1}{h_y} K\left( \frac{y_{i,t} - y}{h_y} \right), 
	\end{aligned}  \right.
\end{equation}
represent Gaussian kernels applied in the $x$ and $y$ directions, respectively. Here, $x_{i,t}$ and $y_{i,t}$ indicate the position coordinates of the $i^{th}$ individual in the $x$- and $y$-directions, respectively, at time $t$, while $N(t)$ is the total number of individuals at that time. The total probability of finding an individual within the observed area is represented by the integral of the PDF function (\ref{PDF_function_2D}) over that area. To calculate the probability of an individual being located in a specific area $A$ with center at $(x,y)$ (or in other words, the percentage of all people in the domain), we integrate the PDF over that area. For sufficiently small areas, the expected number of individuals can be approximated by multiplying the PDF value by the area. This leads to the expression
\begin{equation}
	n(x,y,t) = \hat{f}(x,y,t) \cdot N(t) \cdot A.
\end{equation}
Thus, the smoothed density in the neighborhood of $(x,y)$ at time $t$, is given by the relation 
\begin{equation} \label{KDE_total}
	\rho(x,y,t) = \hat{f}(x,y,t) \cdot N(t).
\end{equation}
Similarly, the two-dimensional PDFs for exposed and susceptible individuals are defined as 
\begin{align} 
	\hat{f}^\text{E}(x, y, t) & = \frac{1}{N^\text{E}(t)} \sum_{i=1}^{N^\text{E}(t)} K_{h_x}\left(x_{i,t}^\text{E} - x\right) K_{h_y}\left(y_{i,t}^\text{E} - y\right), \label{PDF_function_2D_exposed} \\
	\hat{f}^\text{S}(x, y, t) & = \frac{1}{N^\text{S}(t)} \sum_{i=1}^{N^\text{S}(t)} K_{h_x}\left(x_{i,t}^\text{S} - x\right) K_{h_y}\left(y_{i,t}^\text{S} - y\right). \label{PDF_function_2D_susceptible}
\end{align}
Here, $N^\text{E}(t)$ and $N^\text{S}(t)$ represent the total number of exposed and susceptible individuals over the whole stadium at time $t$. The coordinates $x_{i,t}^\text{E}$ and $y_{i,t}^\text{E}$ indicate the position of the $i^{th}$ exposed individual, in the $x-$ and $y-$directions, respectively, while $x_{i,t}^\text{S}$ and $y_{i,t}^\text{S}$ indicate the position of the $i^{th}$ susceptible individual at time $t$. The densities of exposed and susceptible individuals around $(x,y)$ at time $t$, are given respectively by
\begin{align} 
	\rho^\text{E}(x,y,t) & = \hat{f}^\text{E}(x,y,t) \cdot N^\text{E}(t), \label{KDE_exposed} \\
	\rho^\text{S}(x,y,t) & = \hat{f}^\text{S}(x,y,t) \cdot N^\text{S}(t).
\end{align}

\begin{figure} [H]
	\centering
	\includegraphics[width=0.7\linewidth]{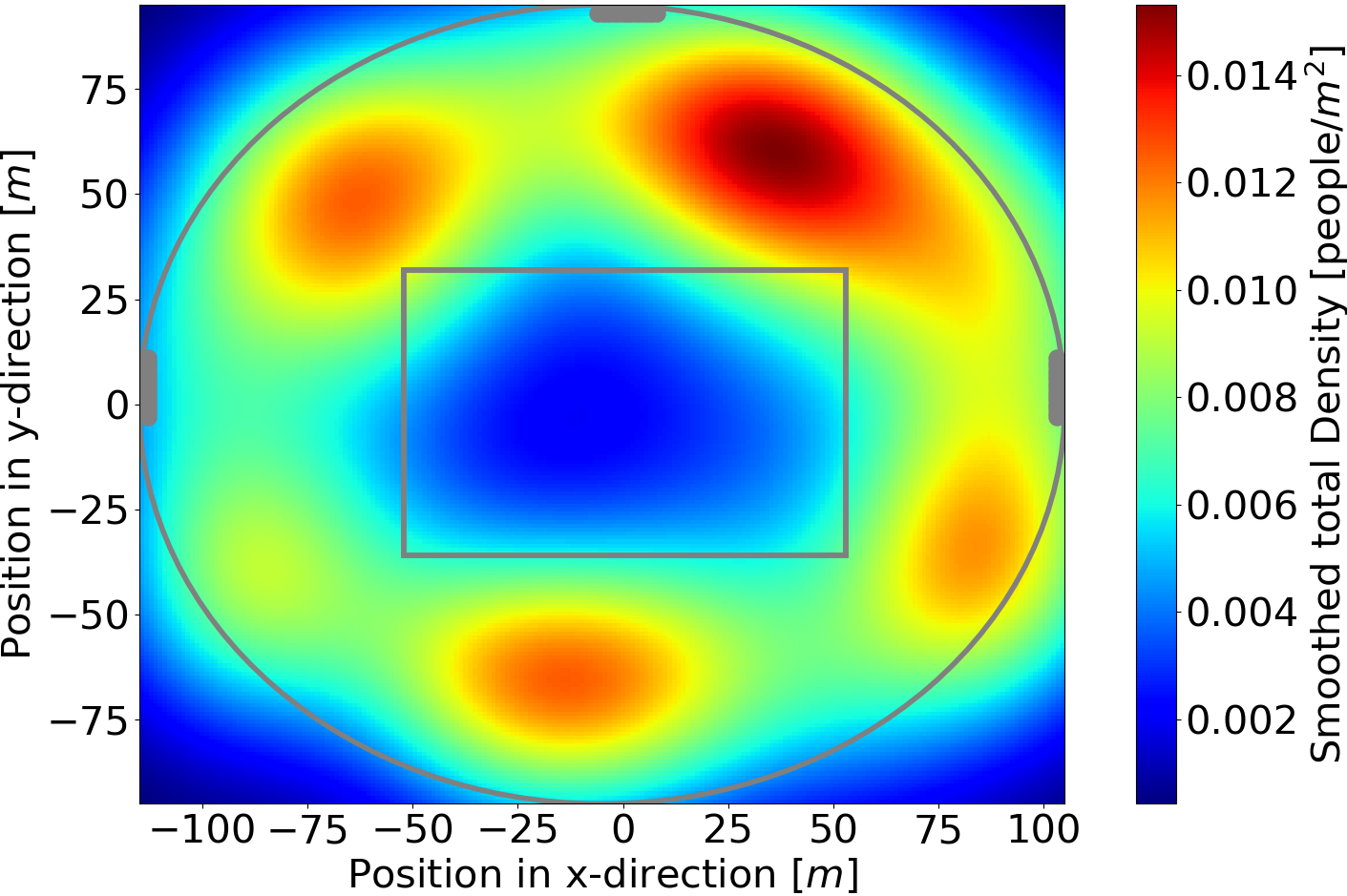}
	\caption{Total smoothed density at the initial time estimated using equation (\ref{KDE_total}).}
	\label{total_density_smooth_901}
\end{figure}
\begin{figure} [H]
	\centering
	\includegraphics[width=0.7\linewidth]{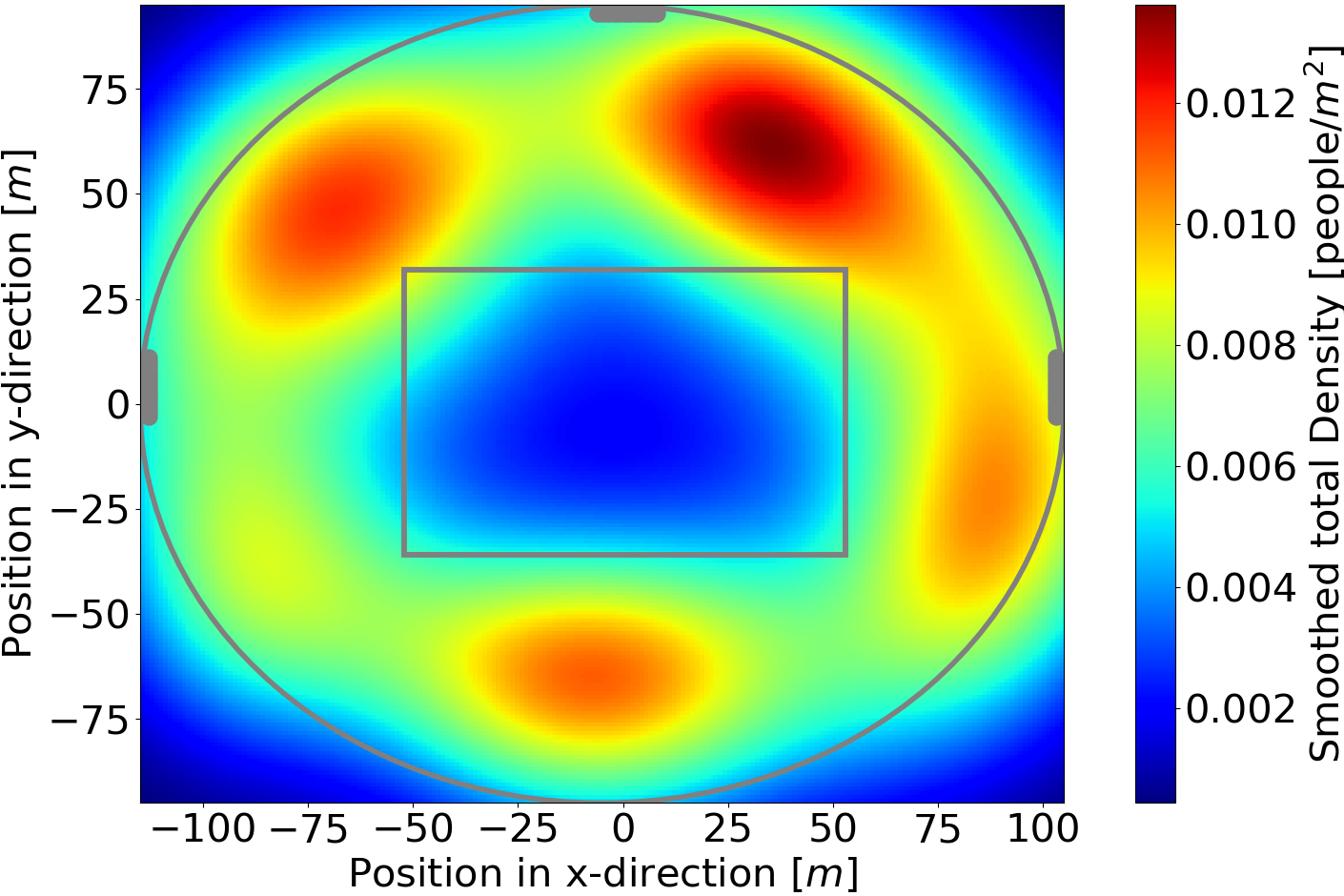}
	\caption{Total smoothed density at $t=450 \text{ s}$ estimated using equation (\ref{KDE_total}).}
	\label{total_density_smooth_945}
\end{figure}
\begin{figure} [H]
	\centering
	\includegraphics[width=0.7\linewidth]{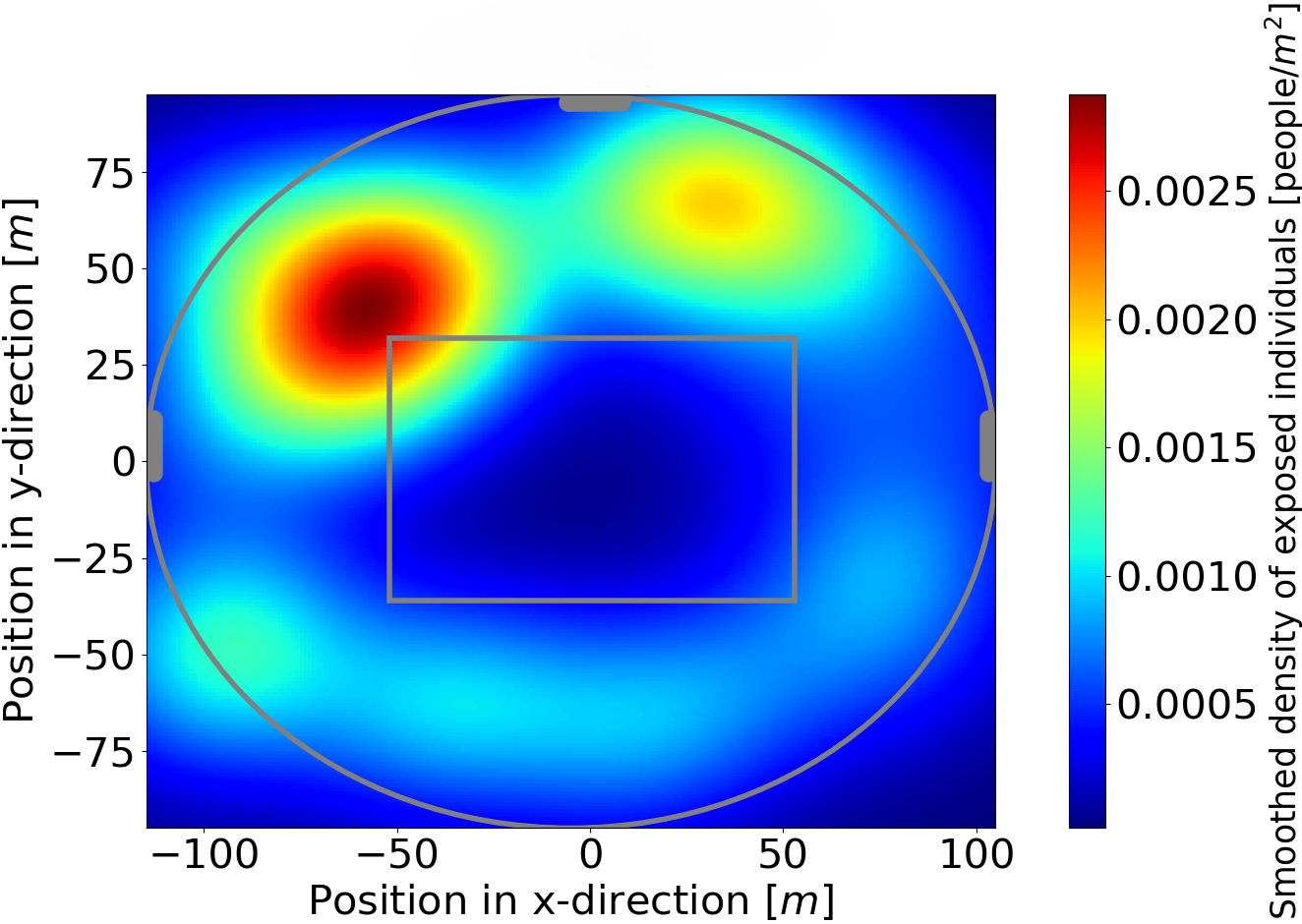}
	\caption{Smoothed density of exposed individuals at $t=450 \text{ s}$ estimated using equation (\ref{KDE_exposed}).}
	\label{exposed_density_smooth_945}
\end{figure}

Figures \ref{total_density_smooth_901}--\ref{exposed_density_smooth_945} show the respective smoothed densities for the time instances considered. We observe that density values are now assigned throughout the domain. We note that the results in Figures \ref{total_density_smooth_901}--\ref{exposed_density_smooth_945} depend on the smoothing method employed, and other choices are also possible, see, e.g., \cite{TreiberMartinKesting}.  

\subsection{Computation of Macroscopic Speeds Magnitudes} 
\begin{figure} [H]
	\centering
	\includegraphics[width=0.7\linewidth]{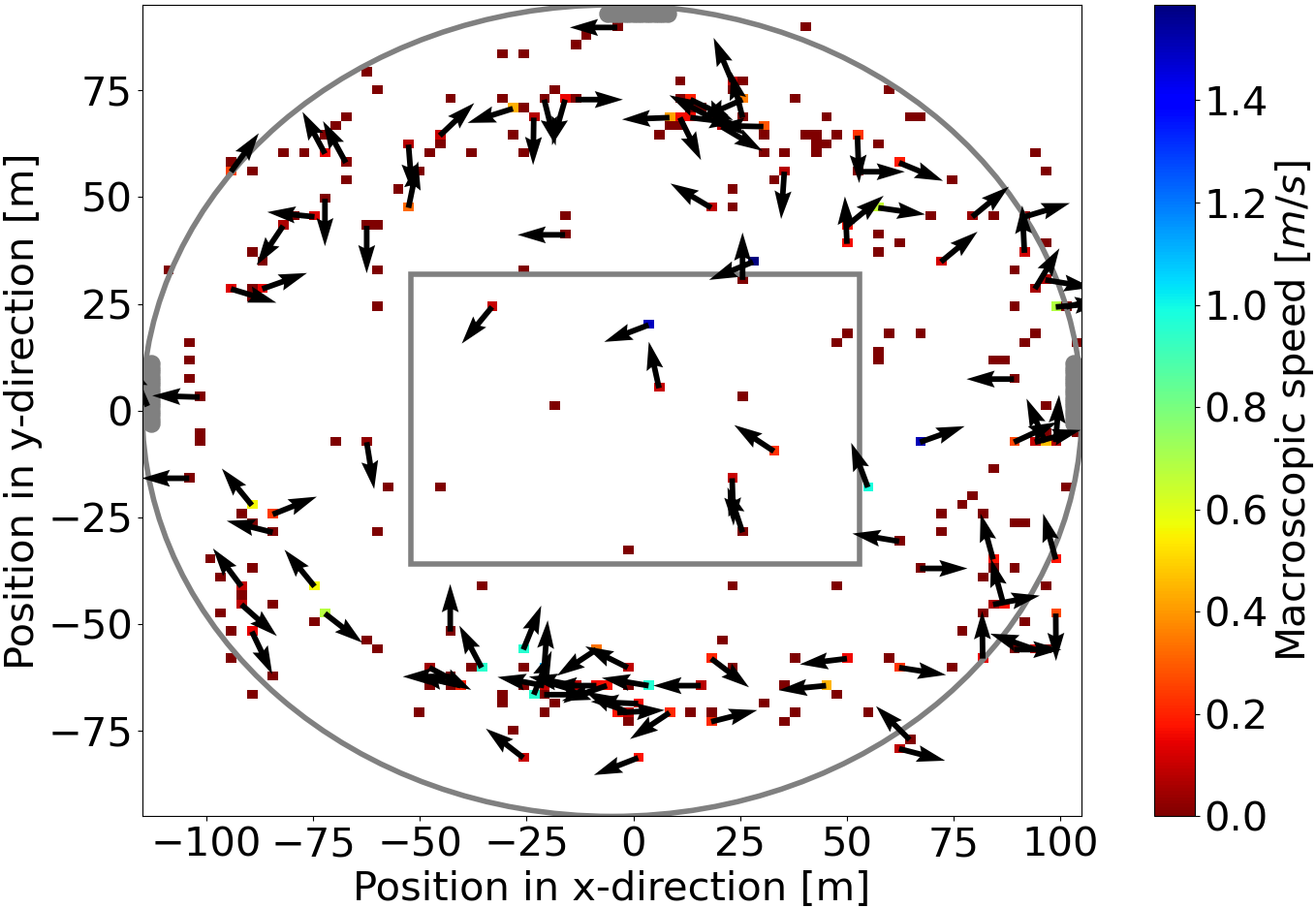}
	\caption{Macroscopic speed at the initial time computed from the data using (\ref{macroscopic_speed}), with direction indicated by arrows, estimated based on angles calculated from equation (\ref{macroscopic_angle_direction}).} 	 
	\label{macroscopic_speed_hist_901}
\end{figure}

\begin{figure} [H]
	\centering
	\includegraphics[width=0.7\linewidth]{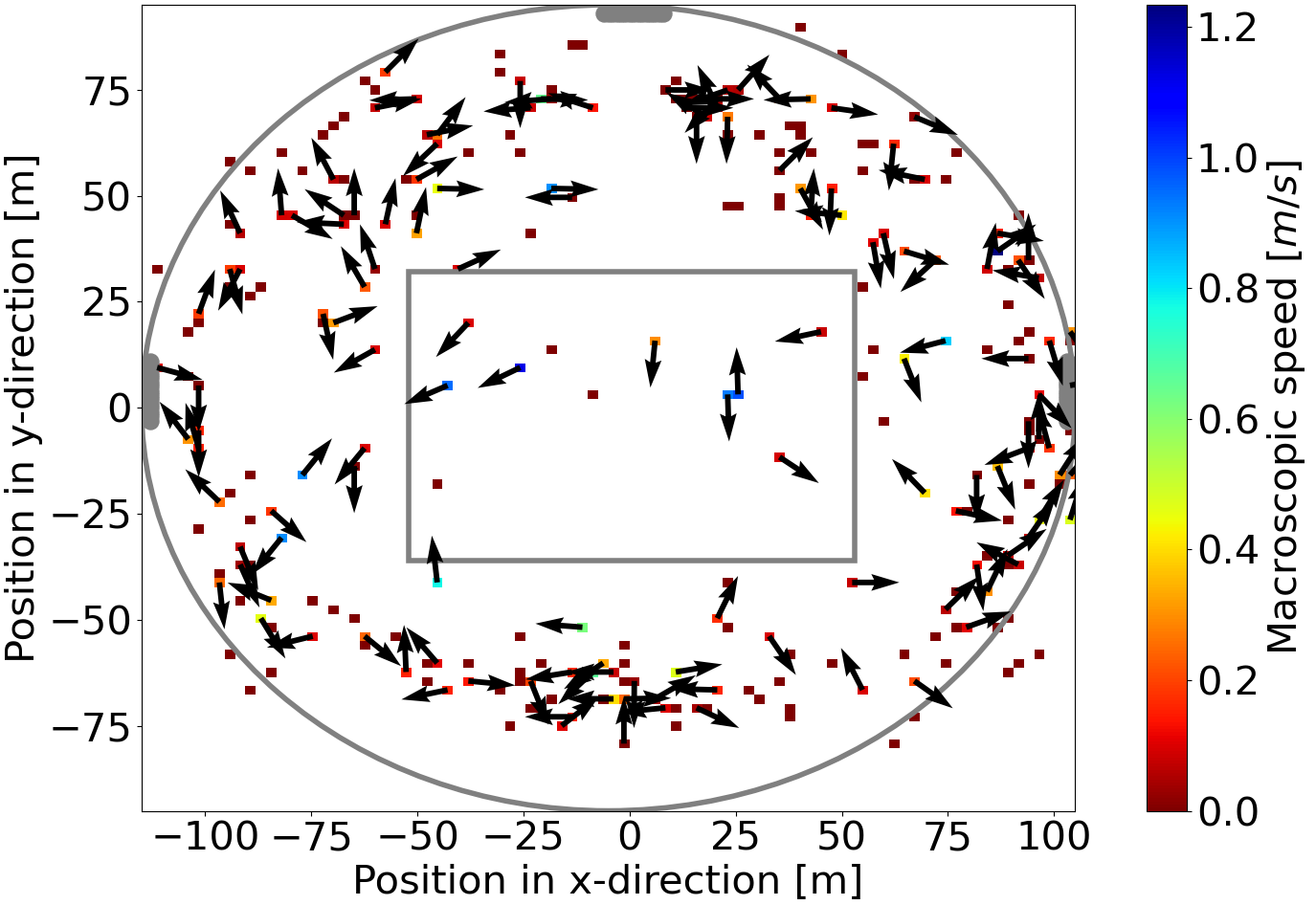}
	\caption{Macroscopic speed at $t=450 \text{ s}$ computed from the data using (\ref{macroscopic_speed}), with direction indicated by arrows, estimated based on angles calculated from equation (\ref{macroscopic_angle_direction}).} 	 
	\label{macroscopic_speed_hist_945}
\end{figure}

To compute the speeds throughout the stadium, we first determine the speed of each individual within each grid square at a specific time instant. The $v_{i}(t)$ and $u_{i}(t)$ represent the speed components in the $x-$ and $y-$directions, respectively, of each individual $i$ at time $t$. 
These components are computed as
\begin{equation} \label{v_u_component_speed}
	v_{i}(t) = \frac{ x_{i,t} - x_{i,t-\Delta t }}{\Delta t}, \quad 	u_{i}(t) = \frac{ y_{i,t} - y_{i,t-\Delta t }}{\Delta t},
\end{equation}  
where $\Delta t = 10$ s denotes the time interval between observations. The magnitude of the speed of each individual $i$ is computed as 
\begin{equation} \label{speed}
	\text{v}_i(t)  = \sqrt{v_{i}^2(t) + u_{i}^2(t)}.
\end{equation}
To compute the magnitude of macroscopic speed $\bar{\text{v}}(x,y,t)$ within a defined region, we discretize the observed area into squares of area $A=5 \text{ m}^2$. For each square, we determine the speed components of all individuals within it using equations (\ref{v_u_component_speed}). Next, we compute the magnitudes of the averaged speed components in the $x-$ and $y-$directions for each square centered at $(x,y)$ as
\begin{equation} \label{u,v_mag}
	\left. \begin{aligned}
		\bar{v}(x,y,t) & = \frac{1}{n(x,y,t)} \sum_{i=1}^{n(x,y,t)} v_i(t), \\ 
		\bar{u}(x,y,t) & = \frac{1}{n(x,y,t)} \sum_{i=1}^{n(x,y,t)} u_i(t),
	\end{aligned} \right.  
\end{equation}
where $n(x,y,t)$ is as defined in Section \ref{Macroscopic_densities}. Then, the magnitude of macroscopic speed for a given square centered at $(x,y)$ is obtained by 
\begin{equation} \label{macroscopic_speed}
	\bar{\text{v}}(x,y,t)  = \sqrt{\bar{v}^2(x,y,t) + \bar{u}^2(x,y,t)}.
\end{equation}
Figures \ref{macroscopic_speed_hist_901} and \ref{macroscopic_speed_hist_945} show the results obtained, where we observe, as in the case of densities computation, the need of apply smoothing to obtain speed values throughout the computational domain. We note that for macroscopic speed computation we do not discern between the type of individuals, as it is reasonably expected that all individuals (on average)  move with the average speed of the crowd. We also note that cells in dark red font and without an arrow correspond to zero speed values. \\

As with density estimation, using interpolation methods we can obtain smoother results, as well as we can assign a speed value throughout the computational domain, even though data may not be available throughout the domain, which is useful towards model calibration. In this case, to estimate macroscopic speed across the observed region, we use a kernel smoothing method, see, e.g., \cite{TreiberMartinKesting}. 

We employ the following Gaussian kernel,
\begin{equation} \label{Gaussian_velocities}
	\hat{K}(x,y) = \frac{1}{2 \pi \sigma^2} e^{ -\left( \frac{ x^2 + y^2 }{2\sigma^2 }\right)},
\end{equation}
where $\sigma$ is positive constant representing the width of the smoothing. The smoothed velocity fields in the $x-$ and $y-$directions, respectively, are given by 
\begin{equation} \label{smoothed_speed_field_x_y_direction}
	\left. \begin{aligned}
		v(x,y,t) & =  \frac{\sum_{i=0}^{N(t)}   e^{ -\left( \frac{ (x-x_{i,t})^2 + (y-y_{i,t})^2 }{2\sigma^2 }\right)}  v_i(t)} { \sum_{i=0}^{N(t)} e^{ -\left( \frac{ (x-x_{i,t})^2 + (y-y_{i,t})^2 }{2\sigma^2 }\right)}}, \\
		u(x,y,t) & =  \frac{\sum_{i=0}^{N(t)}   e^{ -\left( \frac{ (x-x_{i,t})^2 + (y-y_{i,t})^2 }{2\sigma^2 }\right)}  u_i(t)} { \sum_{i=0}^{N(t)}  e^{ -\left( \frac{ (x-x_{i,t})^2 + (y-y_{i,t})^2 }{2\sigma^2 }\right)}}.
	\end{aligned} \right.
\end{equation}
The parameter $\sigma$ is chosen to control the smoothness degree of the velocity field, while ensuring consistency with the actual speed values (see \cite{TreiberMartinKesting} for more details). Here it is set to $\sigma = 5.5 $. Finally, the magnitude of the smoothed velocity field in the observed region is given by 
\begin{equation} \label{Smooth_velocity_field}
	\text{v}(x,y,t) =  \sqrt{v^2(x,y,t) + u^2(x,y,t)},
\end{equation}
where $v(x,y,t)$ and $u(x,y,t)$ are as defined in (\ref{smoothed_speed_field_x_y_direction}). 

We note that the choice of $\sigma$ affects considerably the smoothing results. After conducting several simulation studies we observed that for large values of $\sigma$ the resulting macroscopic speeds are smoother. However, a large $\sigma$ results in a reduction of the speed values, which may not fully reflect the actual, individual speed values from the data. Vice versa, using smaller $\sigma$ values leads to speed values that better align with the observed data, but results in speed values featuring pockets of discontinuities. Thus, here we choose a medium $\sigma$ value to balance between smoothing degree and relevance with the data of the speed values obtained.

\subsection{Computation of Macroscopic Speeds Directions} \label{Macroscopic_speeds_computation_direction}

The direction of movement of the $i^{th}$ individual is defined by the angle between their speed components in the $x-$ and $y-$directions. Each individual has the opportunity to move all over the stadium, so they could have an angle direction between $-\pi$ to $\pi$.
The macroscopic direction of movement in the square, centered at $(x,y)$, at time $t$, is determined by averaging the $v_i(t)$ and $u_i(t)$ components of speed for all individuals $i$ within the square. Thus, the macroscopic angle is 
\begin{align} \label{macroscopic_angle_direction}
	\bar{\theta}(x,y,t) & = \begin{cases} \arctan \left( \frac{\bar{u}(x,y,t)}{\bar{v}(x,y,t)} \right), & \text{if $\bar{v}(x,y,t)>0,$} \\
		\arctan \left( \frac{\bar{u}(x,y,t)}{\bar{v}(x,y,t)}\right) + \pi, & \text{if } \bar{v}(x,y,t)<0, \\ &\text{and } \bar{u}(x,y,t) \geq 0, \\
		\arctan \left( \frac{\bar{u}(x,y,t)}{\bar{v}(x,y,t)}\right) - \pi, & \text{if } \bar{v}(x,y,t)<0, \\ & \text{and } \bar{u}(x,y,t) < 0, \\
		+\frac{\pi}{2}, & \text{if } \bar{v}(x,y,t)=0, \\ & \text{and } \bar{u}(x,y,t)>0, \\
		-\frac{\pi}{2}, & \text{if } \bar{v}(x,y,t)=0, \\ & \text{and } \bar{u}(x,y,t)<0, \end{cases} 
\end{align}
where $\bar{v}(x,y,t)$ and $\bar{u}(x,y,t)$ are defined in (\ref{u,v_mag}).

To estimate the smoothed macroscopic movement direction, we use the smoothed velocity field in the $x-$ and $y-$directions that is given by (\ref{smoothed_speed_field_x_y_direction}). The smoothed macroscopic direction angle $\theta$ is then determined using equation (\ref{macroscopic_angle_direction}), where we use $v$ and $u$ defined by (\ref{smoothed_speed_field_x_y_direction}), instead of $\bar{u}$, $\bar{v}$ defined in (\ref{u,v_mag}).

\begin{figure} [H]
	\centering
	\includegraphics[width=0.7\linewidth]{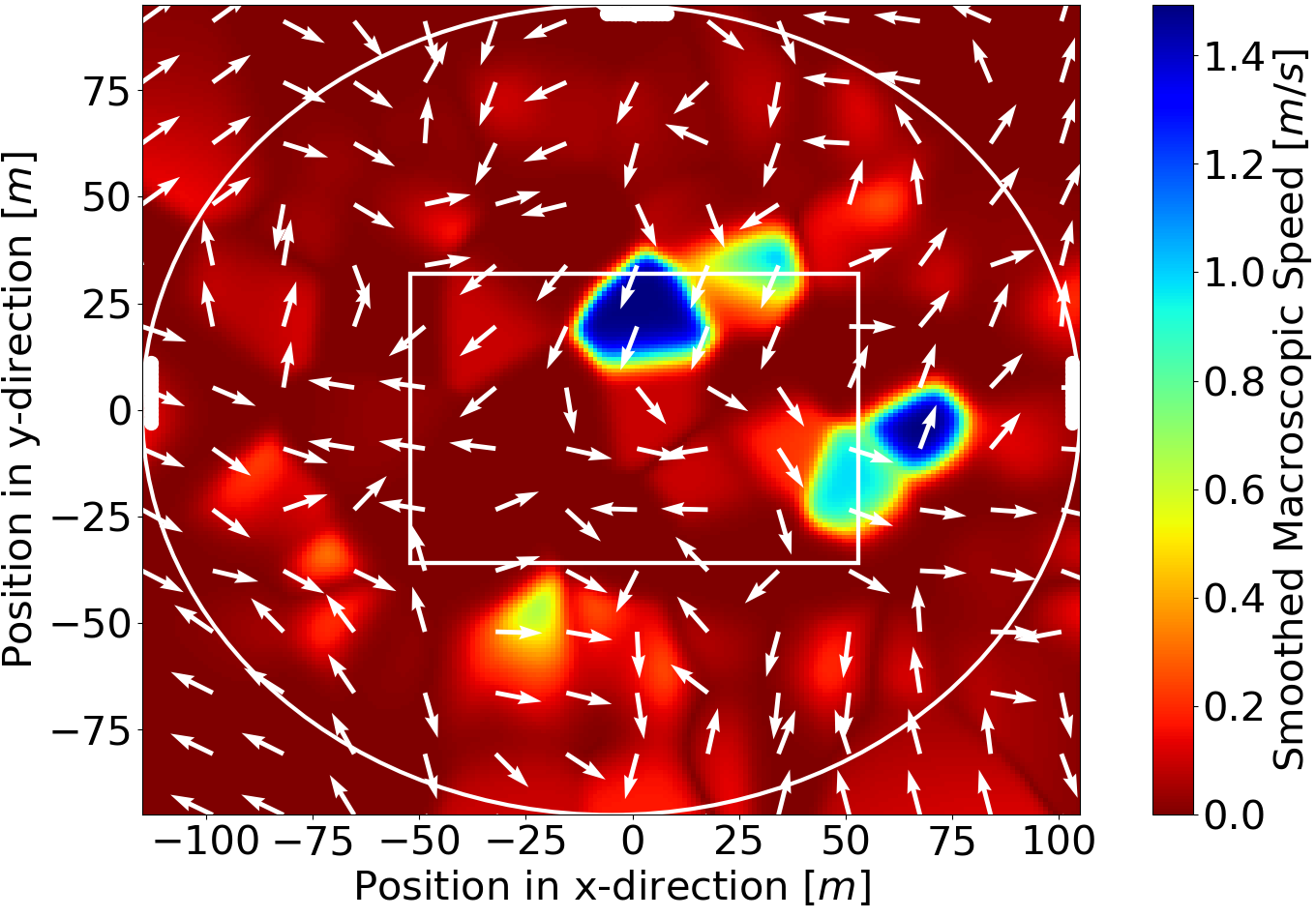}
	\caption{Smoothed macroscopic speed at the initial time estimated using (\ref{Smooth_velocity_field}). White arrows indicate the smoothed direction of movement, computed using (\ref{macroscopic_angle_direction}) with $\bar{v}$ and $\bar{u}$ defined in (\ref{smoothed_speed_field_x_y_direction}).}
	\label{macroscopic_speed_smooth_positions_901}
\end{figure}

\begin{figure} [H]
	\centering
	\includegraphics[width=0.7\linewidth]{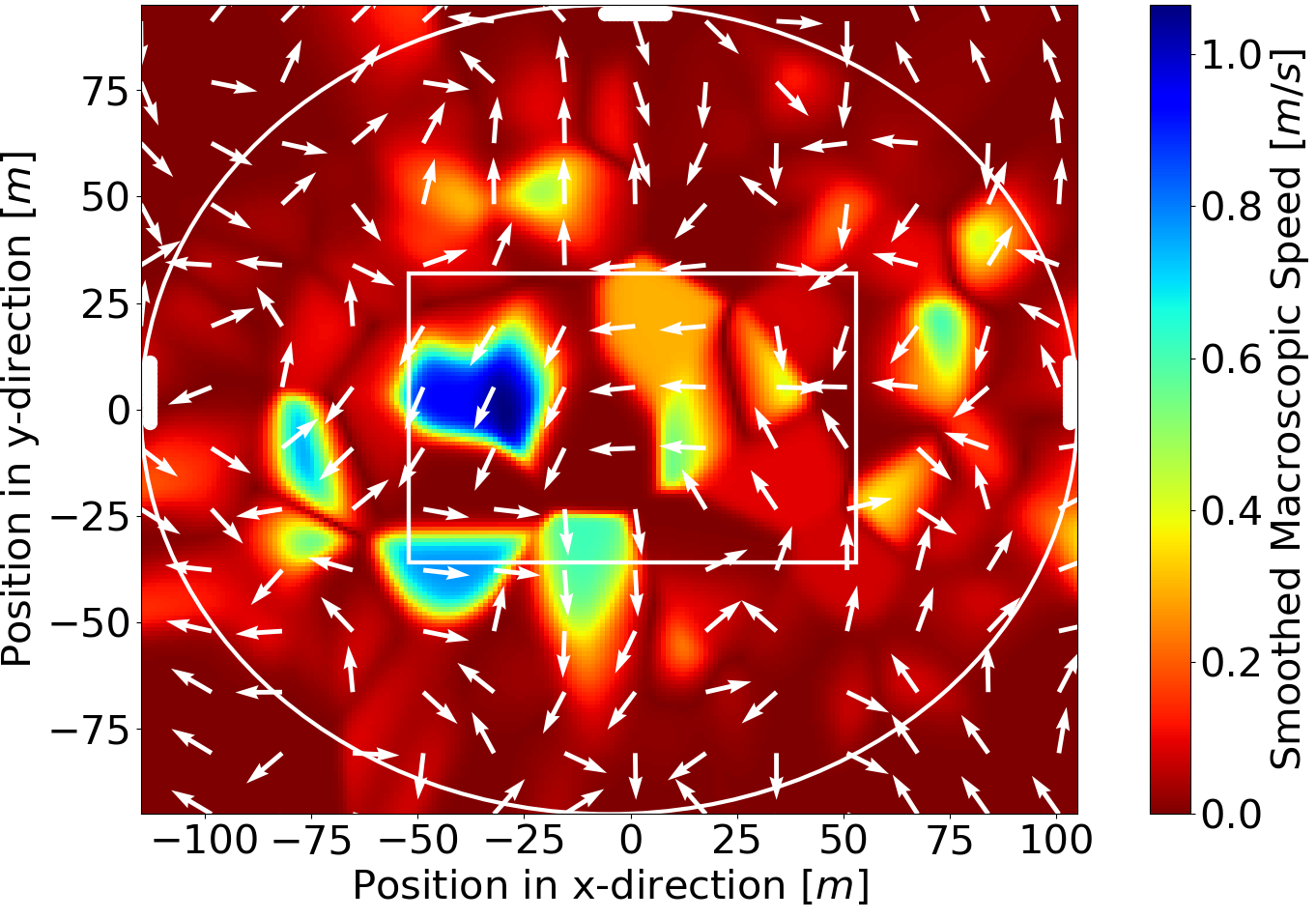}
	\caption{Smoothed macroscopic speed at $t=450 \text{ s}$ estimated using (\ref{Smooth_velocity_field}). White arrows indicate the smoothed direction of movement, computed using (\ref{macroscopic_angle_direction}) with $\bar{v}$ and $\bar{u}$ defined in (\ref{smoothed_speed_field_x_y_direction}).}
	\label{macroscopic_speed_smooth_positions_945}
\end{figure}

Figures \ref{macroscopic_speed_smooth_positions_901} and \ref{macroscopic_speed_smooth_positions_945} display the macroscopic speed across the observed area, estimated using the kernel smoothing method via (\ref{Smooth_velocity_field}). We observe in particular that the directional arrows in the plots of smoothed macroscopic speed, may not always point toward the nearest exits. This can be explained by the fact that computation of the speeds and the directions is based on a sample of individuals in the stadium. As a result, if one exit becomes congested, an individual may change direction of movement toward a less crowded exit, although such a congestion phenomenon is not captured in the available data. In general, the sparse nature of the data (as the actual trajectories data correspond to a fraction of all the people in the stadium) can lead to movement directions that do not accurately reflect the overall crowd movement behavior. Similarly to the density plots, high speed values are observed across the field in both methods, likely due to stadium staff moving freely across the field after the match. 

\section{Coupled Crowd Flow - Epidemic Spreading Model} \label{Model_presentation}
%% Labels are used to cross-reference an item using \ref command.
\begin{table}[H]
	\centering
	\small % Keeps text readable but smaller
	\setlength{\tabcolsep}{3pt} % Adjusts column spacing to fit width
	\renewcommand{\arraystretch}{1.1} % Adjusts row spacing for better readability
	\begin{tabular}{l l l} 
		\textbf{Description}    & \textbf{Notation}  & \textbf{Units}  \\ \hline
		Density of pedestrians  &  $\rho(x,y,t)$ & $\text{ind}/\text{m}^{2}$ \\ 
		Velocity vector with components & $\mathbf{v}(x,y,t)=(v(x,y,t),u(x,y,t))$ & $\text{m}\text{s}^{-1}$  \\
		$v(x,y,t)$ and $u(x,y,t)$  & &  \\  
		Flux flow rate        & $\rho(x,y,t) \mathbf{v}(x,y,t)$ & $\text{m}^{-1} \text{s}^{-1}$ \\  
		Fundamental diagram in $x-$direction  & $V(\rho)=u_\text{max} \exp^{-\alpha (\rho/\rho_\text{max})^2}$  & $\text{m}\text{s}^{-1}$ \\  
		Fundamental diagram in $y-$direction  & $U(\rho)=u_\text{max} \exp^{-\alpha (\rho/\rho_\text{max})^2}$  & $\text{m}\text{s}^{-1}$ \\  
		Dimensionless parameter controlling the  & $\alpha$  &   \\
		decay rate of the speed-density relation & & \\
		Infection rate         & $ \beta_I$         & $\text{s}^{-1}$  \\  
		Direction of motion  & $\boldsymbol{\mu}=\left( \begin{matrix}
			\mu_1 \\
			\mu_2
		\end{matrix} \right)$ & \\  
		Potential from $\Omega$ to destination & $\Phi (\rho)$ & \\  
		Density at the entrance & $\rho_{\text{in}}$ & $\text{ind}/\text{m}^{2}$  \\  
		Velocity at the entrance & $\mathbf{v}_{\text{in}}$ & $\text{m}\text{s}^{-1}$ \\
		Density at the entrance of pedestrian & $\rho^i_{\text{in}}$ & $\text{ind}/\text{m}^{2}$\\  
		group $i$, where $i \in \{\text{S,E,I}\}$  &  & \\  
		Density of pedestrian group $i$,  & $\rho^i(x,y,t)$ & $\text{ind}/\text{m}^{2}$\\  
		at $(x,y,t), $ where $i \in \{\text{S,E,I}\}$  &  & \\  
		Relaxation time & $\tau $ & $\text{s}$ \\  
		Internal pressure function & $P(\rho)=C_0^2 \rho$ & $\text{s}^{-2}$ \\  
		Maximum speed of pedestrians  & $u_\text{max}$ & $\text{m}\text{s}^{-1}$  \\
		Maximum density  & $\rho_\text{max}$ & $\text{ind}/\text{m}^{2}$ \\  
		Anticipation factor & $C_0$ & $\text{m}\text{s}^{-1}$  \\  
		Infection intensity parameter indicating  & $i_0$ &  $\text{m}^{-2} \text{s}^{-1}$\\
		the rate of disease spread per unit area & & \\
		Spatial decay coefficient, indicating & $c_{\boldsymbol{X}}$ & $\text{m}^{-r}$\\
		how infection probability decreases with distance & & \\
		Velocity decay coefficient, indicating how infection & $c_{\boldsymbol{V}}$ & $\text{m}^{-6} \text{s}^{-6}$ \\
		probability decreases with relative speed & & \\
		Exponent scaling the decay rate of infection & $r$ & \\
		 probability with relative distances  & & 
	\end{tabular}
	\caption{Nomenclature of parameters and variables of the model (\ref{crowd_eq_1})--(\ref{function_z}).}
	\label{variable_table}
\end{table}
%\begin{table}[H]
%	\centering
%	\small % Keeps text readable but smaller
%	\setlength{\tabcolsep}{4pt} % Adjusts column spacing to fit width
%	\renewcommand{\arraystretch}{1.2} % Adjusts row spacing for better readability
%	\begin{tabular}{l l l}
%		\textbf{Description}    & \textbf{Notation}  & \textbf{Units}  \\ \hline
%		Characteristic length scale in y-direction & $L_y $ & $m$ \\
%		Characteristic time scale,  $T=\frac{L_x}{u_{\text{max}}}$ and $T=\frac{L_y}{u_{\text{max}}}$  in $x-$   & $T$ & $s$   \\
%		and $y-$direction$ $ \\
%	\end{tabular}
%	\caption{Nomenclature of parameters and variables}
%	\label{variable_table_2}
%\end{table}
%Pedestrians enter from the entrance $\Gamma_\text{in}$ with density $\rho_\text{in}$, and velocity $\mathbf{v}_\text{in}$. %The potential $\Phi$ is equal to zero in $\partial \Omega_E$ since the potential from $(x,y) \in \partial \Omega_E$ to $\partial \Omega_E$ is zero. For convenience, we write the model using $\rho, v,$ and $u$, instead of $\rho(x,y,t), u(x,y,t),$ and $v(x,y,t)$. 
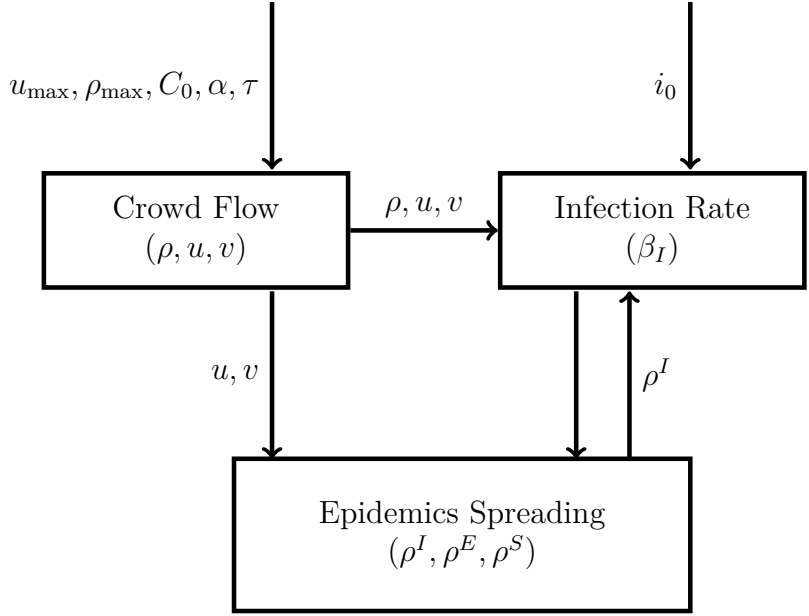
\begin{figure} [H]%[!htb]
	\centering
	\resizebox{0.8\textwidth}{!}{ % Scale to fit within page borders
		\begin{tikzpicture}
			% Draw labeled squares
			\node[draw, line width=0.6mm, minimum width=4cm, minimum height=1.5cm, align=center] 
			(crowd) at (-7.5,-1) {Crowd Flow \\ $(\rho, u, v)$};
			
			\node[draw, line width=0.6mm, minimum width=4cm, minimum height=1.5cm, align=center] 
			(infection) at (-1.5,-1) {Infection Rate \\ $(\beta_I)$};
			
			%\node[draw, line width=0.6mm, minimum width=4cm, minimum height=1.5cm, align=center] 
			%(ventilation) at (4.5,-1) {\textlarger{Ventilation Rate} \\ $(U_G)$};
			\node[draw, line width=0.6mm, minimum width=6cm, minimum height=2cm, align=center] 
			(epidemics) at (-4,-5) {Epidemics Spreading \\ $(\rho^I, \rho^E, \rho^S)$};
			
			% Add arrows
			\draw[line width=0.6mm, ->] (crowd) -- (infection) node[midway, above] {$\rho, u, v$};
			%\draw[line width=0.6mm, ->] (ventilation.west) -- (infection.east);
			
			% Keep the two upper arrows straight
			\draw[line width=0.6mm, ->] (-6.5,2) -- (-6.5,-0.2) node[midway, left] {$ u_\text{max}, \rho_\text{max}, C_0, \alpha, \tau$};
			\draw[line width=0.6mm, ->] (-1,2) -- (-1,-0.2) node[midway, left] {$ i_0$};
			
			%\draw[line width=0.6mm, ->] (4.5,-3) -- (ventilation.south) node[pos=-0.6, above] {Input velocity field (control variable)};
			\draw[line width=0.6mm, ->] (-6.5,-1.8) -- (-6.5,-4) node[midway, left] {$u,v$};
			
			% Restore original two lower arrows
			\draw[line width=0.6mm, ->] (-2.5,-1.8) -- (-2.5,-4);
			\draw[line width=0.6mm, ->] (-1.8,-4) -- (-1.8,-1.8) node[midway, right] {$\rho^I$};
			
		\end{tikzpicture}
	}
	\caption{Graphical representation of the coupled crowd–epidemic modeling framework, showing the interaction between crowd flow and infection dynamics.}
	\label{model_figure}
\end{figure}
The crowd flow model is depicted in Figure \ref{model_figure} and is given by (see, e.g., \cite{DiseaseContagion}, \cite{delis} for details)
\begin{align}
	\rho_t + (\rho v)_x + (\rho u)_y & = 0 \label{crowd_eq_1}, &\\
	(\rho v)_t + (v^2 \rho + C_0^2 \rho )_x + (uv \rho )_y & = \frac{1}{\tau} \rho (V(\rho)\mu_1 - v),& \label{momentum_eq_x}  \\
	(\rho u)_t + (uv \rho )_x +(u^2 \rho + C_0^2 \rho )_y & = \frac{1}{\tau} \rho (U(\rho)\mu_2  - u),&  \label{momentum_eq_y}  \\
	\rho (x,y,0) & = \rho_0(x,y),  & \\
	v(x,y,0) & = v_0(x,y), & \\
	u(x,y,0) & = u_0(x,y), & \\
	\rho (x,y,t) & = \rho_{\text{in}}(x,y,t),  \qquad \qquad \forall (x,y) \in \Gamma_\text{in},  &  \label{boundary_condition_1}\\
	\nabla \rho(x,y,t) \cdot \mathbf{n} &= 0, \qquad \qquad \qquad  \forall (x,y) \in \Gamma_\text{W} \cup \Gamma_\text{E}, &  \\
	\mathbf{v}(x,y,t) &= \mathbf{v}_\text{in}(x,y,t),  \qquad \qquad  \forall (x,y) \in \Gamma_\text{in}, & \\
	\mathbf{v}(x,y,t) \cdot \mathbf{n} = 0, & \frac{\partial}{\partial \mathbf{n}}(\mathbf{v}(x,y,t) \cdot \mathbf{l}) = 0,  \quad \forall (x,y) \in \Gamma_\text{W}, & \\
	%\mathbf{v}(x,y,t) &= -(\mathbf{v}(x,y,t) \cdot \mathbf{n})\mathbf{n} + (\mathbf{v}(x,y,t) \cdot \mathbf{l}) \mathbf{l}, & \forall (x,y) \in \Gamma_\text{W}, \\
	\mathbf{v}(x,y,t)  &= u_\text{max} \mathbf{n},  \quad \qquad \qquad \forall (x,y) \in \Gamma_\text{E}, & \label{crowd_flow_model_last_eq} 
\end{align}
where $(x,y) \in \Omega \subset \mathbb{R}^2$ is a bounded domain with boundary $\partial \Omega = \Gamma_{\text{W}} \cup \Gamma_{\text{E}} \cup \Gamma_\text{in}$, representing the walls, the exits, and the entrances, respectively. In the boundary conditions, we also denote by $\mathbf{n}$ and $\mathbf{l}$ the normal and the tangent vectors at the boundary, respectively. The initial conditions for the states are denoted by $(\rho_0, v_0, u_0)$, while the remaining parameters and variables are defined in Table \ref{variable_table}.\\

To obtain the boundary conditions (\ref{boundary_condition_1})--(\ref{crowd_flow_model_last_eq}), we make the following assumptions. We assume that once an individual reaches the boundary walls, they move along the wall (free-slip boundary conditions) to the nearest exit. Furthermore, we assume that individuals exit the domain at maximum speed. The internal pressure function $P(\rho) = C_0^2 \rho$ in the momentum equations (\ref{momentum_eq_x}) and (\ref{momentum_eq_y}), models repulsive interactions between pedestrians and generates forces that drive motion from high-density to low-density regions.  The terms $ \frac{1}{\tau} \rho (V(\rho) \mu_1 - v)$,  $\frac{1}{\tau} \rho (U(\rho) \mu_2 - u)$ 
represent the relaxation terms and show how fast pedestrians modify their speed according to the desired one, namely, $V(\rho)$ and $U(\rho)$ in $x-$direction and $y-$direction, respectively. The desired direction of motion $\boldsymbol{\mu}$ is defined assuming that each pedestrian knows the overall density distribution of the crowd and that each pedestrian moves at the direction that minimizes travel time to an exit (see e.g., \cite{PedestrianDynamics}, \cite{Cristiani2014}). Thus, pedestrians move as
\begin{equation}
	\boldsymbol{\mu}=\left( \begin{matrix}
		\mu_1 \\
		\mu_2
	\end{matrix} \right) = - \frac{1}{\Vert \nabla \Phi(\rho) \Vert} \nabla \Phi(\rho),
\end{equation} 
where the potential $\Phi : \Omega \rightarrow \mathbb{R}$ is defined by the Eikonal equation (see, e.g., \cite{DiseaseContagion})
\begin{align} 
	\Vert \nabla \Phi(\rho) \Vert &= C(\rho), &(x,y) \in \Omega, \label{Eikonal_eq1} \\
	\Phi(\rho) &= 0, &(x,y) \in \Gamma_{\text{E}},  \label{Eikonal_eq2} 
\end{align}
where $C(\rho)$ is a running cost function, which is inversely proportional to desired speeds, i.e., 
\begin{equation}
	C(\rho)= \left( \begin{matrix}
	1/{V(\rho)} \\
	1/{U(\rho)}
\end{matrix} \right).
\end{equation} 
%\begin{figure}[h] 
%	\begin{center}
	%		\includegraphics[width=0.4\textwidth]{V_exp}
	%		\caption{Speed-density relation} \label{V_exp}
	%	\end{center}
%\end{figure}
%The second choice based on the fact that pedestrians aim at following the shortest path to the destination based on the memory of its location, and temper their behavior locally to avoid high densities. This leads to the following definition of $\mu_1$, $\mu_2$,
%\begin{align}
%	& \mu_1 = \frac{-\nabla \Phi(\rho) - \omega \nabla [V(\rho)^{-1} + g(\rho)]}{\Vert - \nabla \Phi(\rho) - \omega \nabla [V(\rho)^{-1} + g(\rho)] \Vert}, \\
%	& \mu_2 = \frac{-\nabla \Phi(\rho) - \omega \nabla [U(\rho)^{-1} + g(\rho)]}{\Vert - \nabla \Phi(\rho) - \omega \nabla [U(\rho)^{-1} + g(\rho)] \Vert}.			
%\end{align}
%\vspace{5mm}\\
%In this case of $\mu$, we have the alternative definition of the speed-density relations $V(\rho), U(\rho)$:
%\begin{align}
%	& V(\rho) = u_\text{max} (1-\rho/\rho_\text{max}), \\
%	& U(\rho) = u_\text{max} (1-\rho/\rho_\text{max}).
%\end{align}
The crowd flow model is coupled with the epidemic spreading model (see \cite{DiseaseContagion}, \cite{delis})
\begin{align} 
	\rho^\text{S}_t + (\rho^\text{S} v)_x + (\rho^\text{S}u)_y & = -\beta_\text{I} \rho^\text{S}, &  \label{contagion_model_1} \\ 
	\rho^\text{E}_t + (\rho^\text{E} v)_x + (\rho^\text{E}u)_y & = \beta_\text{I} \rho^\text{S}, &\\
	\rho^\text{I}_t + (\rho^\text{I} v)_x + (\rho^\text{I}u)_y & = 0 \label{rho_I}, &\\
	\rho^\text{S}_0 & = \lambda \rho_0, & \label{theoritical_boundary_epidemic_1}\\
	\rho^\text{I}_0 & = (1-\lambda) \rho_0, & \label{theoritical_boundary_epidemic_2} \\
	\rho^\text{E}_0 & = 0,  & \label{boundary_condition_epidemic_1}\\
	\rho^l(x,y,t) &= \rho^l_{\text{in}}(x,y,t), & \forall \left( x,y \right) \in \Gamma_{\text{in}} \label{boundary_condition_epidemic_2} \\
	%\rho^\text{E}(x,y,t) &= 0, & \forall \left( x,y \right) \in \Gamma_{\text{in}} \\
	%\rho^\text{S}(x,y,t) &= 0, & \forall \left( x,y \right) \in \Gamma_{\text{in}} \\ %\rho^\text{S}_\text{in}
	\nabla \rho^l(x,y,t) \cdot \mathbf{n} &= 0 , & \forall \left(x,y \right) \in \Gamma_\text{W} \cup \Gamma_\text{E}, \label{contagion_model_2} 
\end{align}
where $l \in \{ \text{S}, \text{E}, \text{I}\}$ denotes the type of individual being susceptible (S), exposed (E), or infected (I), and $\lambda$ is a constant number, denoting the initial percentage of susceptible individuals. We note that the total density satisfies $\rho = \rho^\text{S} + \rho^\text{I} + \rho^\text{E}$. The infection rate $\beta_I$ of the epidemic spreading model (\ref{contagion_model_1})--(\ref{contagion_model_2}) is based on social interaction and is defined as (see \cite{DiseaseContagion}) 
\begin{equation} \label{beta_I}
	\beta_I(\boldsymbol{x}) =\scalebox{1.5}{$\displaystyle \int_{\Omega}$} z \left(\boldsymbol{x}-\boldsymbol{y}, \mathbf{v}\boldsymbol{(x)}-\mathbf{v}\boldsymbol{(y)} \right) \frac{\rho^I(\boldsymbol{y})}{\rho(\boldsymbol{y})} d\boldsymbol{y},
\end{equation} 
where $\boldsymbol{x} = (x_1, y_1)$, $\boldsymbol{y}=(x_2,y_2)$, and 
\begin{equation} \label{function_z}
	z(\boldsymbol{x}, \mathbf{v}(\boldsymbol{x})) = i_0z_{\boldsymbol{X}}(\boldsymbol{x})z_{\boldsymbol{V}}(\mathbf{v}(\boldsymbol{x})),
\end{equation}
with $z_{\boldsymbol{X}}(\boldsymbol{x})= e^{-c_{\boldsymbol{X}}\Vert \boldsymbol{x}\Vert^r}$, which indicates that infections depend on the distance between the pedestrians, $z_{\boldsymbol{V}}(\mathbf{v}(\boldsymbol{x}))= e^{-c_{\boldsymbol{V}}\Vert \mathbf{v}(\boldsymbol{x})\Vert^6}$, which indicates that infections depend on their relative velocities, and
$i_0$ is determined by the infectivity, which depends on high-level considerations about the spreading severity, such as, e.g., on whether pedestrians wear masks or whether there is the effect of ventilation (see \cite{DiseaseContagion}, \cite{delis}).

\section{Numerical Scheme for the Crowd-Epidemic Transport Model} \label{Numerical_scheme_presentation}
\subsection{Numerical Solution of the Crowd Flow Model}
For the numerical simulations, we use a finite volume scheme. We assume a spatial domain of dimensions $\Omega = [0, L_x] \times [0,L_y]$, and divide the solution space $\Omega$ into a uniform computational grid. The grid spacing in the $x-$ and $y-$directions are denoted by $\Delta x$ and $\Delta y$, respectively, and the time step by $\Delta t$.  We write the equations (\ref{crowd_eq_1})--(\ref{momentum_eq_y}) in conservative form as
\begin{equation} \label{conservative_form}
	\mathbf{Q}_t + \mathbf{F}(\mathbf{Q})_x + \mathbf{G}(\mathbf{Q})_y = \mathbf{S}(\mathbf{Q}),
\end{equation} 
where $\mathbf{Q} = (\rho, \rho v, \rho u )^T$ are the conservative variables, $\mathbf{F}$ and $\mathbf{G}$ are the fluxes in $x-$ and $y-$directions, respectively, and $\mathbf{S}$ is the source term. They are given by
\begin{equation} \label{terms_of_consevative_form}
	\mathbf{F}(\mathbf{Q}) = \left( \begin{matrix}
		\rho v \\
		\rho v^2 + C_0^2 \rho \\
		\rho v u 
	\end{matrix} \right), \mathbf{G}(\mathbf{Q}) = \left( \begin{matrix}
		\rho u \\
		\rho v u \\
		\rho u^2 + C_0^2 \rho
	\end{matrix}\right), \mathbf{S} = \left( \begin{matrix}
		0 \\
		\frac{1}{\tau} \rho \left( -V(\rho) \mu_1 -v \right) \\
		\frac{1}{\tau} \rho \left( -U(\rho) \mu_2 -u \right) 
	\end{matrix} \right).
\end{equation} 
\noindent The final finite volume scheme is given by
\begin{equation}
	\mathbf{Q}_{i,j}^{n+1} = \mathbf{Q}_{i,j}^n - \frac{\Delta t}{\Delta x} \left( \mathbf{F}_{i+\frac{1}{2}, j}^n - \mathbf{F}^n_{i-\frac{1}{2},j} \right) - \frac{\Delta t}{\Delta y} \left( \mathbf{G}^n_{i,j+\frac{1}{2}} - \mathbf{G}^n_{i,j-\frac{1}{2}} \right)  + \Delta t \mathbf{S}_{i,j}^n,
\end{equation}
where  $i,j$ is the index pair for $\left( x_i, y_j \right)$. For the numerical fluxes in both directions, i.e., for $\mathbf{F}$ and $\mathbf{G}$, respectively, we use the Rusanov scheme, as described in \cite{leveque2002finite}, which is given by
\begin{equation}
	\mathbf{F}^n_{i-\frac{1}{2},j} = \frac{1}{2} \left[ \mathbf{F} \left(\mathbf{Q}^n_{i-1,j} \right) + \mathbf{F} \left(\mathbf{Q}^n_{i,j} \right) - \delta_{i-\frac{1}{2},j} \left(\mathbf{Q}^n_{i,j}-\mathbf{Q}^n_{i-1,j}\right) \right],
\end{equation}
\begin{equation}
	\mathbf{G}^n_{i,j-\frac{1}{2}} = \frac{1}{2} \left[ \mathbf{G} \left(\mathbf{Q}^n_{i,j-1} \right) + \mathbf{G} \left(\mathbf{Q}^n_{i,j} \right) - \delta_{i,j-\frac{1}{2}} \left(\mathbf{Q}^n_{i,j}-\mathbf{Q}^n_{i,j-1} \right) \right],
\end{equation}
where $\delta_{i-\frac{1}{2},j}$ and $\delta_{i,j-\frac{1}{2}}$ are defined as
\begin{align}
	\delta_{i-\frac{1}{2},j} &= \max \left( \left|  \frac{\partial\mathbf{F}}{\partial\mathbf{Q}} \right|_{\mathbf{Q} = \mathbf{Q}^n_{i-1,j}} , \left| \frac{\partial\mathbf{F}}{\partial\mathbf{Q}}\right|_{\mathbf{Q} = \mathbf{Q}^n_{i,j}} \right), \\
	\delta_{i,j-\frac{1}{2}} &= \max \left( \left|  \frac{\partial\mathbf{G}}{\partial\mathbf{Q}} \right|_{\mathbf{Q} = \mathbf{Q}^n_{i,j-1}} , \left| \frac{\partial\mathbf{G}}{\partial\mathbf{Q}}\right|_{\mathbf{Q} = \mathbf{Q}^n_{i,j}} \right).
\end{align}
For the boundary conditions, we assume a layer of ghost cells surrounding the computational domain. At the walls, we impose free-slip conditions, at the exits, we enforce the maximum velocity, and at the entries, we prescribe an inflow velocity, coincident with the theoretical boundary conditions (\ref{boundary_condition_1})--(\ref{crowd_flow_model_last_eq}) (see also, e.g., \cite{on_epidemic_spreading_mitigation_via_heuristic_macro_control_of_people_flow}), that is,
\begin{align}
	\rho_g^n &= \rho^n_{i,j}, \quad &\mathbf{v}^n_g &= \mathbf{v}^n_{i,j} - 2 \left(\mathbf{v}^n_{i,j} \cdot \mathbf{n} \right) \mathbf{n}, \quad & \forall \left(x_i,y_j \right) \in \Gamma_\text{W}, \\
	\rho_g^n &= \rho^n_{i,j}, \quad &\mathbf{v}^n_g &= u_{\text{max}} \mathbf{n}, \quad  & \forall \left(x_i,y_j \right) \in \Gamma_\text{E}, \\
	\rho_g^n &= \rho^n_{\text{in}, i, j}, \quad & \mathbf{v}^n_g &= \mathbf{v}^n_{\text{in}, i, j}, & \forall \left(x_i,y_j \right) \in \Gamma_\text{in}, \label{numerical_boundary_condition_crowd}
\end{align}
where $\mathbf{v}_g$, $\rho_g$ represent the velocity and density at the ghost cells. \\

\subsection{Numerical Computation of the Direction Vector}
We numerically obtain the direction vector $\boldsymbol{\mu}$ in the source term by solving the Eikonal equation (\ref{Eikonal_eq1}), (\ref{Eikonal_eq2}) at each time step, using the Fast Sweeping Method (FSM) as detailed in \cite{Fast_Sweeping_Method}, \cite{FadiKaldawi}. A brief presentation of this method is given next.
To compute the numerical solution of the Eikonal equation, we use a Godunov upwind difference scheme to discretize the PDE equation at interior grid points, which is given by
\begin{align} \label{FSM_equation}
	\left[ \left(\Phi^n_{i,j} - \min \left( \Phi^n_{i-1,j}, \Phi^n_{i+1,j} \right) \right)^+ \right]^2 &+ \left[ \left( \Phi^n_{i,j} - \min \left( \Phi^n_{i,j-1}, \Phi^n_{i,j+1} \right) \right)^+ \right]^2& \\
	&= \left( h c_{i,j}^n \right)^2, & \nonumber \\
	\Phi^n_{i,j} &= 
	\begin{cases}
		0, & \forall (x_i, y_j) \in \Gamma_{\text{E}}, \\
		10^{12}, & \forall (x_i, y_j) \in \Omega \setminus \Gamma_{\text{E}},
	\end{cases}
\end{align}
where 
\begin{equation}
	(x)^+ = 
	\begin{cases}
		x, & x > 0, \\
		0, & x \leq 0,
	\end{cases}
\end{equation}
and $i = 1, \ldots, N_x,$ $j = 1, \ldots,  N_y$. The parameters $N_x$ and $N_y$ are the number of gridpoints in the $x-$ and $y-$directions respectively. We note that $c^n_{i,j}=C(Q_{i,j}^n)$, where $C(\cdot)$ denotes the cost function as defined in \cite{FadiKaldawi}. The parameter $h$ represents the spatial step where here we choose $h = \Delta x = \Delta y$.  Equation (\ref{FSM_equation}) is applied for a `sweep' from the left side to the right and from the bottom to the top of the domain. A `sweep' refers to a systematic pass through the computational grid in a specific direction to update the solution values. Each sweep follows a fixed order and uses information from neighboring points to update the current point. We apply `sweeps' in four directions, defined as 
\begin{align}
	& \quad  i = 1, \ldots, N_x, \quad j=1, \ldots, N_y, \\
	& \quad  i = N_x, \ldots, 1,\quad j = 1, \ldots, N_y, \\
	& \quad  i = N_x, \ldots, 1,\quad j = N_y, \ldots, 1, \\
	& \quad  i = 1, \ldots, N_x,\quad j = N_y, \ldots, 1.
\end{align}
The process for each $n$ ends when the following convergence condition is satisfied
\begin{equation} \label{condition_FSM}
	\Vert \Phi^n - \Phi^{n, \text{new}} \Vert_2 < \delta,
\end{equation}
for some convergence threshold $\delta > 0$, where $\Vert . \Vert_2$ denotes the $L^2$-norm over the spatial grid. The values of $\Phi^{n, \text{new}}$ at each time step $n$ are obtained after completing the four directional sweeps. During each sweep, the value $\Phi^{n, \text{new}}_{i,j}$ of $\Phi^{n, \text{new}}$ is computed as,
\begin{equation}
	\Phi^{n, \text{new}}_{i,j} = \min \left( \Phi^n_{i,j}, \bar{\Phi}^n_{i,j} \right),
\end{equation}
where $\Phi^n_{i,j}$ is the value of $\Phi^n$ at the point $(x_i, y_j)$ from the previous sweep, i.e. the current estimate. The term $\bar{\Phi}^n_{i,j}$ is computed by solving the local update equation,
\begin{equation} 
	\bar{\Phi}^n_{i,j} = 
	\begin{cases}
		\min \left( \left( \Phi_{i,j}^{n} \right)^x, \left( \Phi_{i,j}^{n} \right)^y \right) + c_{i,j}h, \quad \left| \left( \Phi_{i,j}^{n} \right)^x - \left( \Phi_{i,j}^{n} \right)^y \right| \ge c_{i,j}h, \\
		\left( \Phi_{i,j}^{n} \right)^x + \left( \Phi_{i,j}^{n} \right)^y + \sqrt{2c_{i,j}^2 h^2 - \left(\left( \Phi_{i,j}^{n} \right)^x - \left( \Phi_{i,j}^{n} \right)^y \right)^2}, \quad \text{otherwise}.
	\end{cases}
\end{equation}
Here, the directional components are defined as:
\begin{equation}
	\left( \Phi_{i,j}^{n} \right)^x = \min \left( \Phi^n_{i+1,j}, \Phi^n_{i-1,j}\right),  \quad	\left( \Phi_{i,j}^{n} \right)^y =  \min \left( \Phi^n_{i,j+1}, \Phi^n_{i,j-1}\right).
\end{equation}
Once the convergence condition (\ref{condition_FSM}) is satisfied, the values in $\Phi^n_\text{new}$ represent the solution to equation (\ref{FSM_equation}). \\

\subsection{Numerical Solution of the Epidemic Spreading Model}
We employ a finite-volume scheme, similar to that used for the crowd flow model, to approximate the solution of the contagion model (\ref{contagion_model_1})--(\ref{contagion_model_2}). We express equations (\ref{contagion_model_1})--(\ref{contagion_model_2}) in conservative form as follows,
\begin{equation} \label{conservative_form_contagion}
	\mathbf{W}_t + \mathbf{L}(\mathbf{W})_x + \mathbf{M}(\mathbf{W})_y = \mathbf{T}(\mathbf{W}),
\end{equation} 
where $\mathbf{W} = (\rho^\text{S}, \rho^\text{E}, \rho^\text{I} )^T$ are the conservative variables, $\mathbf{L}$ and $\mathbf{M}$ are the flux vectors in $x-$ and $y-$directions, respectively, and $\mathbf{T}$ is the source term. They are given by
\begin{equation} \label{terms_of_consevative_form_contagion}
	\mathbf{L}(\mathbf{W}) = \left( \begin{matrix}
		\rho^\text{S} v \\
		\rho^\text{E} v \\
		\rho^\text{I} v
	\end{matrix} \right), \quad \mathbf{M}(\mathbf{W}) = \left( \begin{matrix}
		\rho^\text{S} u \\
		\rho^\text{E} u \\
		\rho^\text{I} u
	\end{matrix}\right), \quad \mathbf{T} = \left( \begin{matrix}
		- \beta_\text{I} \rho^\text{S} \\
		\beta_\text{I} \rho^\text{S} \\
		0
	\end{matrix} \right).
\end{equation}
The finite volume scheme is given by
\begin{equation}
	\mathbf{L}_{i,j}^{n+1} = \mathbf{L}_{i,j}^n - \frac{\Delta t}{\Delta x} \left( \mathbf{L}_{i+\frac{1}{2}, j}^n - \mathbf{L}^n_{i-\frac{1}{2},j} \right) - \frac{\Delta t}{\Delta y} \left( \mathbf{M}^n_{i,j+\frac{1}{2}} - \mathbf{M}^n_{i,j-\frac{1}{2}} \right)  + \Delta t \mathbf{T}_{i,j}^n.
\end{equation}
\noindent For the numerical fluxes in both directions, we employ the Rusanov scheme, similarly to the crowd flow model, which is given by
\begin{align}
	\mathbf{L}^n_{i-\frac{1}{2},j} &= \frac{1}{2} \left[ \mathbf{L} \left(\mathbf{W}^n_{i-1,j} \right) + \mathbf{L} \left(\mathbf{W}^n_{i,j} \right) - \eta_{i-\frac{1}{2},j} \left(\mathbf{W}^n_{i,j}-\mathbf{W}^n_{i-1,j}\right) \right], \\
	\mathbf{M}^n_{i,j-\frac{1}{2}} &= \frac{1}{2} \left[ \mathbf{M} \left(\mathbf{W}^n_{i,j-1} \right) + \mathbf{M} \left(\mathbf{W}^n_{i,j} \right) - \eta_{i,j-\frac{1}{2}} \left(\mathbf{W}^n_{i,j}-\mathbf{W}^n_{i,j-1} \right) \right],
\end{align}
where $\eta_{i-\frac{1}{2},j}$ and $\eta_{i,j-\frac{1}{2}}$ are defined as
\begin{align}
	\eta_{i-\frac{1}{2},j} &= \max \left( \left|  \frac{\partial\mathbf{L}}{\partial\mathbf{W}} \right|_{\mathbf{W} = \mathbf{W}^n_{i-1,j}} , \left| \frac{\partial\mathbf{L}}{\partial\mathbf{W}}\right|_{\mathbf{W} = \mathbf{W}^n_{i,j}} \right), \\
	\eta_{i,j-\frac{1}{2}} &= \max \left( \left|  \frac{\partial\mathbf{M}}{\partial\mathbf{W}} \right|_{\mathbf{W} = \mathbf{W}^n_{i,j-1}} , \left| \frac{\partial\mathbf{M}}{\partial\mathbf{W}}\right|_{\mathbf{W} = \mathbf{W}^n_{i,j}} \right).
\end{align}
The infection rate is computed by numerically integrating (\ref{beta_I}) over the computational domain as %\textcolor{red}{Diorthosi}
\begin{equation} \label{beta_I_numerically}
	\beta^n_{I,i,j} = \sum_{k=0}^{N_x-1} \sum_{d=0}^{N_y-1} z^n_{i,j,k,d} \frac{\text{W}_{k,d}^{n,3}}{\mathbf{Q}_{k,d}^{n,1}} \Delta x \Delta y, %\sum_{m=1}^{3} \text{W}_{k,d}^{n,m}
\end{equation}
where $\mathbf{Q}_{k,d}^{n,1}$ denotes the first component of $\mathbf{Q}$ which corresponds to the total density. The numerical form of (\ref{function_z}) becomes %\text{W}^{m} denotes the $m-$th components of $\mathbf{W}$
\begin{equation}
	z^n_{i,j,k,d}= i_0 z_{\boldsymbol{X}}(\boldsymbol{x}_{i,j} - \boldsymbol{x}_{k,d}) z_{\boldsymbol{V}}(\mathbf{v}^n_{i,j} - \mathbf{v}^n_{k,d}),
\end{equation}
where 
\begin{align}
	z_{\boldsymbol{X}}(\boldsymbol{x}_{i,j} - \boldsymbol{x}_{k,d};\boldsymbol{x}_{i,j} ) &= \frac{e^{-c_{\boldsymbol{X}} \Vert \boldsymbol{x}_{i,j} - \boldsymbol{x}_{k,d}\Vert^r}}{\sum_{m=0}^{N_x-1} \sum_{n=0}^{N_y-1} e^{-c_{\boldsymbol{X}} \Vert \boldsymbol{x}_{i,j} - \boldsymbol{x}_{m,n}\Vert^r}}, \label{z_x}\\
	z_{\boldsymbol{V}}(\mathbf{v}^n_{i,j} - \mathbf{v}^n_{k,d};\mathbf{v}^n_{i,j} ) &= \frac{e^{-c_{\boldsymbol{V}} \Vert \mathbf{v}^n_{i,j} - \mathbf{v}^n_{k,d}\Vert^6}}{\sum_{m=0}^{N_x-1} \sum_{n=0}^{N_y-1} e^{-c_{\boldsymbol{V}} \Vert \mathbf{v}^n_{i,j} - \mathbf{v}^n_{m,n}\Vert^6}}. \label{z_v}
\end{align}
%gia na eimaste consistent me to numerically implementation aytou toy paper
For consistency with the numerical implementation of the model as described in \cite{DiseaseContagion}, the parameters $c_{\boldsymbol{X}}$ and $c_{\boldsymbol{V}}$ are set equal to one. In addition, the equations (\ref{z_x}) and (\ref{z_v}) are normalized such that
\begin{equation}
	\sum_{k=0}^{N_x-1} \sum_{d=0}^{N_y-1} z_{\boldsymbol{X}}(\boldsymbol{x}_{i,j} - \boldsymbol{x}_{k,d};\boldsymbol{x}_{i,j}) = \sum_{k=0}^{N_x-1} \sum_{d=0}^{N_y-1} z_{\boldsymbol{V}}(\mathbf{v}^n_{i,j} - \mathbf{v}^n_{k,d};\mathbf{v}^n_{i,j})  = 1
\end{equation}
%\begin{equation}
%	\int_{\Omega} z_{\boldsymbol{X}}( \boldsymbol{x}) d \boldsymbol{x} = \int_{\Omega}z_{\boldsymbol{V}}(\mathbf{v(\boldsymbol{x})}) d\boldsymbol{x} = 1.
%\end{equation} 
Similarly to the boundary conditions of the crowd flow model, at the domain edges we employ ghost cells. Their values, for each $l \in \{\text{S}, \text{E}, \text{I}\}$, are assigned as
\begin{align}
	\rho_g^{l, n} &= \rho^{l, n}_{i,j},  &\forall \left(x_i,y_j \right) \in \Gamma_\text{W} \cup \Gamma_\text{E}, \\
	 \rho_g^{I, n} = (1-\lambda) \rho^n_{\text{in},i,j} , &\quad  \rho_g^{S, n} = \lambda \rho^n_{\text{in},i,j},  & \forall \left(x_i,y_j \right) \in \Gamma_\text{in}, \label{numerical_boundary_conditions_epidemic}
\end{align}
where the choice (\ref{numerical_boundary_conditions_epidemic}) is made for consistency with (\ref{theoritical_boundary_epidemic_1}), (\ref{theoritical_boundary_epidemic_2}), and (\ref{numerical_boundary_condition_crowd}).
%\section*{Part II: Computation of Macroscopic Densities and Velocities}

\section{Model Calibration} \label{Model_calibration}
\subsection{Calibration of the Crowd Flow Component with Data 13 Minutes After the Match Ended} \label{Crowd_flow_calibration}
We start calibrating the crowd flow component of the model, as it constitutes the main element of the complete model. For the calibration, we use the Nelder-Mead algorithm \cite{NelderMead} (see also  \cite{SPILIOPOULOU} for calibration of traffic flow models in particular). The model is simulated from $t=0\text{ s}$ to $T=200 \text{ s}$, with a time step of $\Delta t=0.1 \text{ s}$. The chosen time interval starts $13$ minutes after the match ended, while within the time period of $200$ s considered, no pedestrian enters or leaves the stadium. The spatial domain has dimensions $220\text{ m} \times 190\text{ m} $ and is discretized with a grid spacing of $\Delta x=1.7\text{ m}$ and $\Delta y = 1.7\text{ m}$, resulting in $N_x \times N_y$ grid points in the $x-$ and $y-$directions, respectively, where $N_x=129$ and $N_y=111$. We fix the parameters $u_{\text{max}}=1.4\text{ m}\text{s}^{-1}$ and $\rho_{\text{max}}=6 \text{ ind}/\text{m}^2 $ (see \cite{delis}). In the two-dimensional case, the algorithm operates on a simplex defined by three vertices in the plane. Each vertex corresponds to a set of model parameters $\mathbf{p} = [\alpha, C_0, \tau]$. For each parameter set, the density within the domain predicted by the model (\ref{crowd_eq_1})--(\ref{crowd_flow_model_last_eq})  is compared to the density obtained from the data. This comparison is performed through a cost function, defined as the root mean squared error (RMSE) between the predicted and observed densities,
\begin{equation} \label{cost_function}
	J(\mathbf{p})= \sqrt{\frac{1}{K N_x N_y} \sum_{k=0}^{K-1} \sum_{i=0}^{N_x-1} \sum_{j=0}^{N_y-1}  \left( b_{\text{pred}}(i,j,k; \mathbf{p}) - b_{\text{obs}}(i,j,k) \right)^2 },
\end{equation}
%\begin{equation} \label{cost_function}
%	J(\mathbf{p})= \sqrt{\frac{1}{K} \sum_{k=0}^{K-1} \left[ \sum_{i=0}^{N_x} \sum_{j=0}^{N_y}  \left( b_{\text{pred}}(i,j,k; \mathbf{p}) - b_{\text{obs}}(i,j,k; \mathbf{p}) \right) \Delta x \Delta y \right]^2},
%\end{equation}
 where $b_{\text{pred}}(i,j,k; \mathbf{p})=\rho(x_i,y_j,t_k; \mathbf{p})$ is the model-predicted densities with parameters $\mathbf{p}$ within the computational domain at the discrete time $t_k$, with $t_k = \frac{10k}{ \Delta t}$ for $k=0, \dots, K-1$. The term $b_{\text{obs}}(i,j,k) =\rho_{\text{obs}}(x_i,y_j,k)$ contains the corresponding observed densities from the data computed in Section \ref{ComputationOfMacroscopicDensitiesAndSpeeds}. %We choose for model calibration the density data from areas close to the three exits because at that locations the data are in general richer, as there is a higher degree of movement.
 
 For each evaluation of the cost function, the numerical solution of the model is computed using the current parameter values $\mathbf{p}=(\alpha,C_0,\tau)$, yielding the predicted density $b_{\text{pred}}(i,j,k; \mathbf{p})=\rho(x_i,y_j,t_k; \mathbf{p})$. The index $k$ runs over all time instances, with $K$ denoting the total number of time instances for which data are available. At each iteration, the Nelder-Mead algorithm updates the simplex by adjusting its vertices to reduce the cost function. The algorithm stops once one of the following two criteria is satisfied. The first criterion is that all simplex vertices are sufficiently close to the vertex with the lowest value of the cost function, i.e.,
\begin{equation} \label{1st_criterion}
	\max_{i \in \{0,1,2\}} \Vert \mathbf{p}_i - \mathbf{p}^{\text{lowest}} \Vert_{\infty}< 0.1,
\end{equation}
where $\mathbf{p}_i$ is the $i-th$ simplex vertex, and $\mathbf{p}^{\text{lowest}}$ is the vertex with the lowest cost function value in (\ref{cost_function}). This criterion ensures that all parameters have stabilized to within a tolerance of $0.1$. This value was determined empirically based on preliminary tests to balance convergence accuracy and computational cost. The second criterion is that a maximum of $20$ iterations is reached, which prevents excessive computation.  

\begin{figure} [H]
	\centering
	\includegraphics[width=0.8\linewidth]{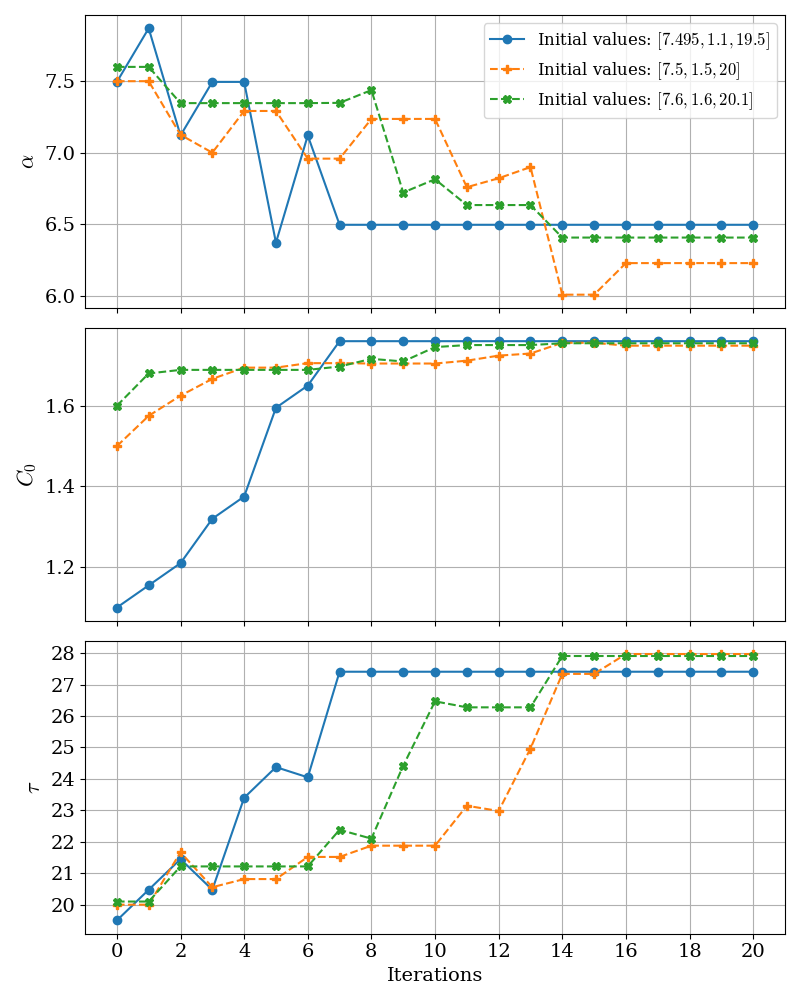}
	\caption{Convergence of the calibrated parameters $\alpha$, $C_0$, and $\tau$, starting from different initial value sets.}
	\label{parameter_convergence}
\end{figure}

\begin{figure} [H]
	\centering
	\includegraphics[width=0.65\linewidth]{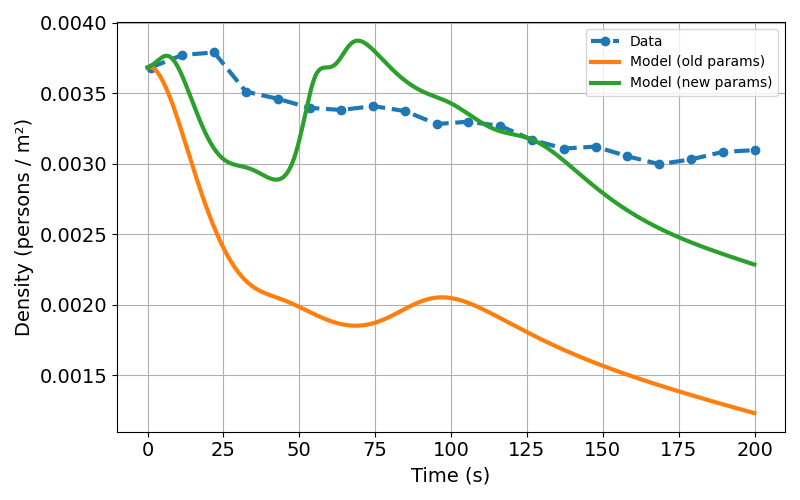}	
	\caption{Time evolution of density at the spatial location $(-82.8, 65.3)$. Observed data are compared with model predictions obtained using the initial parameter set, $\mathbf{p}_{\text{init}} = [7.5, 1.5, 20]$, and the calibrated parameter set which is $\mathbf{\bar{p}} = [6.23, 1.75, 27.97]$. The corresponding relative error (\ref{relative_error}) is $23.9\%$.}
	\label{calibration_density_-95_-67}
\end{figure}

\begin{figure} [H]
	\centering
	\includegraphics[width=0.65\linewidth]{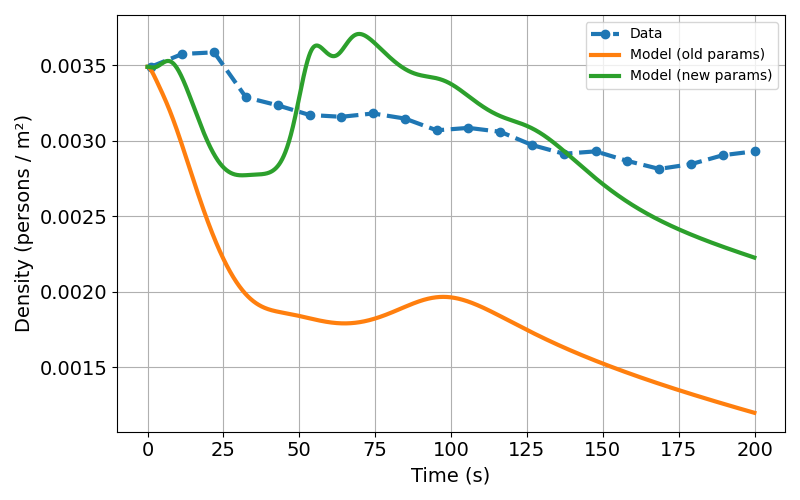}
	
	\caption{Time evolution of density at the spatial location $(-84.4, 63.7)$. Observed data are compared with model predictions obtained using the initial parameter set, $\mathbf{p}_{\text{init}} = [7.5, 1.5, 20]$, and the calibrated parameter set which is $\mathbf{\bar{p}} = [6.23, 1.75, 27.97]$. The corresponding relative error (\ref{relative_error}) is $21.7\%$.}
	\label{calibration_density_-95_69}
\end{figure}

To assess the robustness of the calibration, we perform multiple repetitions starting from different initial guesses $\mathbf{p}_{\text{init}} = [\alpha_{\text{init}}, C_{0,\text{init}}, \tau_{\text{init}}]$ and verify whether the estimated parameters converge to similar values. As illustrated in Figure \ref{parameter_convergence}, when initialized from different, but closely spaced initial conditions, the calibrated parameters converge to almost identical values. Minor deviations are due to numerical approximation errors. 
We then evaluate the performance of the calibration approach. We choose three specific spatial points to visualize the calibration results, based on the minimum, supremum norm (over time). Specifically, for each spatial location $(x_i, y_j)$, we define an error as
\begin{equation} \label{choice_of_points}
	E(x_i,y_j) = \max_{k \in \{0, \dots K-1\}} \vert \bar{\rho}(x_i, y_j, t_k;\mathbf{\bar{p}}) - \rho_{\text{obs}}(x_i, y_j, k) \vert,
\end{equation}
where $\bar{\rho}$ denotes the model-predicted densities with the optimal parameters, say, $\mathbf{\bar{p}}$, corresponding to the values obtained in the calibration. Based on this error, we identify the three spatial points $\{ (x_{i_l}, y_{i_l})\}^3_{l=1}$ corresponding to the three smallest values of $E(x_i,y_j)$, for all $(x_i,y_j) \in \Omega' $, where $\Omega'$ is the descritized domain. To quantify the agreement between observed and predicted data, we evaluate the relative error, i.e., 
\begin{equation} {\label{relative_error}}
	\varepsilon_{\mathrm{rel}}(x_i,y_j)=\max_{k \in \{0, \dots K-1\}} \frac{\left| \bar{\rho}(x_i, y_j, t_k;\mathbf{\bar{p}}) - \rho_{\text{obs}}(x_i, y_j, k)\right|}{ \rho_{\text{obs}}(x_i, y_j, k)} .
\end{equation}%In Figs. \ref{alpha_less}--\ref{tau_less}, the convergence of the calibrated parameters is shown for the initial value set $\mathbf{p}_{\text{init}} = [3, 0.2, 0.3]$. %As shown, the parameters $C_0$ and $\tau$ converge to stable values after approximately 50 iterations, while $\alpha$ increases without bound. The same behavior is observed for the initial value set $\mathbf{p}_{\text{init}} = [10, 0.8, 0.9]$, as shown in Figs. \ref{alpha_high}--\ref{tau_high}. In contrast, the calibration using the initial parameter set $\mathbf{p}_{\text{init}} = [7.5, 0.5, 0.6]$ converges to stable values for all calibrated parameters, as shown in Figs. \ref{alpha_baseline}--\ref{tau_baseline}. This indicates that the convergence of $\alpha$ is sensitive to the choice of initial values. 
\begin{figure} [H]
	\centering
	\includegraphics[width=0.65\linewidth]{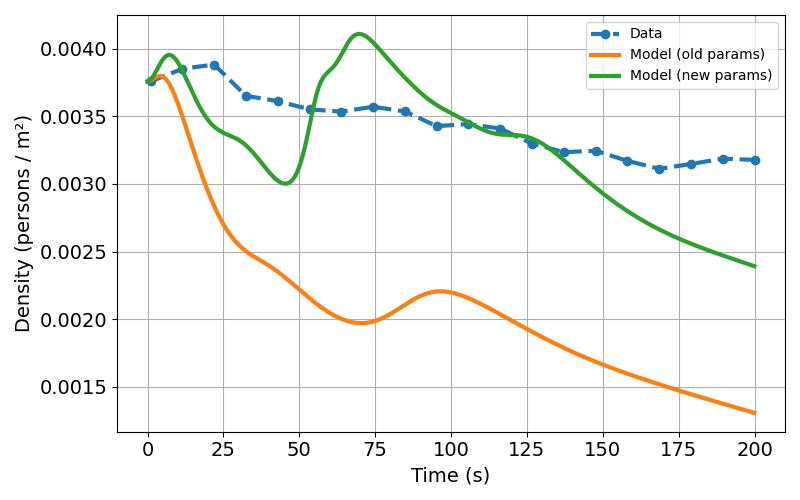}
	\caption{Time evolution of density at the spatial location $(-81.1, 68.3)$. Observed data are compared with model predictions obtained using the initial parameter set, $\mathbf{p}_{\text{init}} = [7.5, 1.5, 20]$, and the calibrated parameter set which is $\mathbf{\bar{p}} = [6.23, 1.75, 27.97]$. The corresponding relative error (\ref{relative_error}) is $22.4\%$.}
	\label{calibration_density_-95_70}
\end{figure}
Figures \ref{calibration_density_-95_70}--\ref{calibration_density_-95_69} show the time evolution of density, at these three discrete spatial points, where the observed data are compared with the model predictions obtained using both the initial and the calibrated parameter sets. The observed total densities are on the order of $0.001 \text{ ped}/ \text{m}^2$, since the calibration is performed using the smoothed data and the sparsity of the available observations results in density values of this magnitude. %The time window corresponds to about one hour after the end of the match, and thus, the total densities obseved are in the range of $0.001 \text{ ped}/ \text{m}^2$. 
We also observe that the model matches the data more accurately when calibrated parameters are used, as expected. However, the agreement may not be consistenlty accurate across all spatial locations, which may be attributed to the fact that the smoothed densities are derived from a sparse dataset. The relative error corresponding to Figures \ref{calibration_density_-95_70}--\ref{calibration_density_-95_69} is approximately $23\%$, which may be deemed appropriate given the sparsity of the available data and the smoothing procedure used to estimate the densities. 

%\begin{figure} [H]
%	\centering
%	\includegraphics[width=0.6\linewidth]{tau_baseline.png}
%	\label{tau_baseline}
%	\caption{Convergence of the parameter $\tau$ starting from the initial value set $\mathbf{p}_{\text{init}} = [7.5, 0.5, 0.6]$. The plot shows the evolution over iterations of  $\tau$.}
%\end{figure}
%\begin{figure} [H]
%	\centering
%	\subfloat[\label{alpha_high}]{\includegraphics[width=0.6\linewidth]{alpha_high.png}}
%	\subfloat[\label{C0_high}]{\includegraphics[width=0.6\linewidth]{C0_high.png}}
%	\vspace{0.5cm}
%	\subfloat[\label{tau_high}]
%	{\includegraphics[width=0.6\linewidth]{tau_high.png}}
%	\caption{Convergence of the parameters starting from the initial value set $\mathbf{p}_{\text{init}} = [10, 0.8, 0.9]$. Subplots show the evolution over iterations of (\ref{alpha_high}) $\alpha$, (\ref{C0_high}) $C_0$, (\ref{tau_high}) $\tau$.}
%\end{figure}

\subsection{Calibration of the Crowd Flow Component with Data 50 minutes After the Match Ended} \label{crowd_flow_calibration_50_mins_after}
To validate our calibration approach also in a different scenario, as well as in cases where constraints have to be incorporated in the Nelder-Mead algorithm, we present here additional calibration results. In this case, the crowd flow component is calibrated using a different time interval, namely, starting $50$ minutes after the end of the match, because we observed by exhaustive simulations that this window also resulted in satisfactory calibration of the epidemic component of the model, which is presented in the next subsection. The parameter bounds are selected based on empirical obsevations from several calibration runs performed over different time windows. 
\begin{figure} [H]
	\centering
	\includegraphics[width=0.7\linewidth]{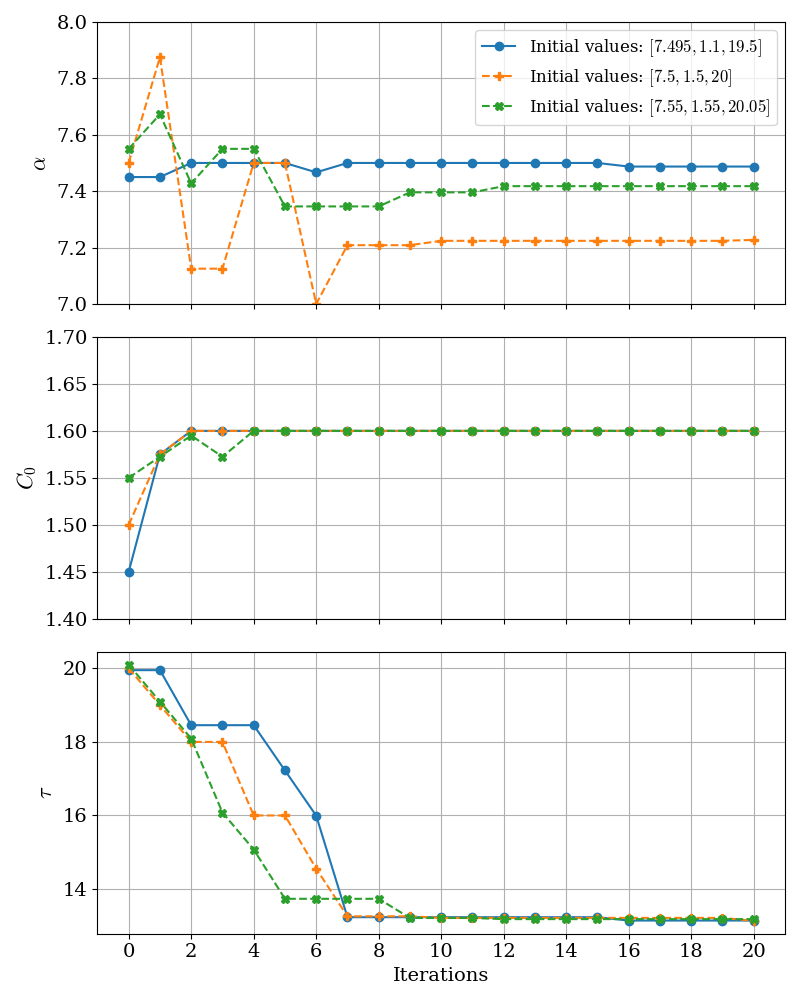}
	\caption{Convergence of the calibrated parameters $\alpha$, $C_0$, and $\tau$, starting from different initial value sets.}
	\label{parameter_convergence_50_mins_after}
\end{figure}
Following the same calibration procedure as in Section \ref{Crowd_flow_calibration}, the only difference is that the parameters $\alpha, C_0,$ and $\tau$ are constrained to the intervals
\begin{equation}
	\alpha \in [6, 8], \qquad C_0 \in [1.4, 1.6], \qquad \tau \in [12, 20.5].
\end{equation}  
The robustness of the calibration is assessed by performing multiple optimization runs from different initial parameter guesses. The corresponding results are shown in Figure  \ref{parameter_convergence_50_mins_after}. We observe in particular that the calibrated value of $C_0$ converges to the upper bound of the constraint set, while the constraints of the other parameters are not activated.

\begin{figure} [H]
	\centering
	\includegraphics[width=0.7\linewidth]{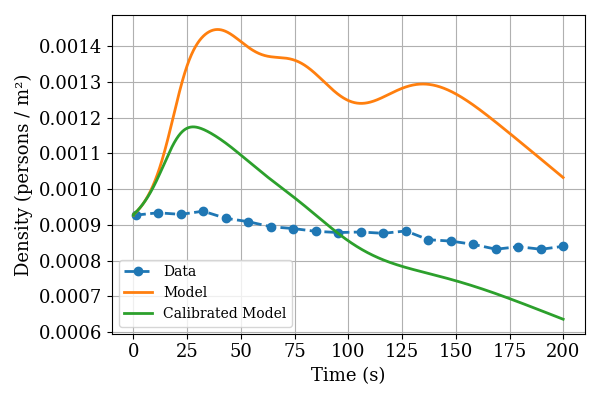}
	
	\caption{Time evolution of density at the spatial location $(-95.3, 70.4)$. Observed data are compared with model predictions obtained using the initial parameter set and the calibrated parameter set which is $\mathbf{\tilde{p}} = [7.23, 1.6, 13.13]$. The corresponding relative error (\ref{relative_error}) is $24.9\%$.}
	\label{total_density_-95.281_70.369}
\end{figure}

\begin{figure} [H]
	\centering
	\includegraphics[width=0.7\linewidth]{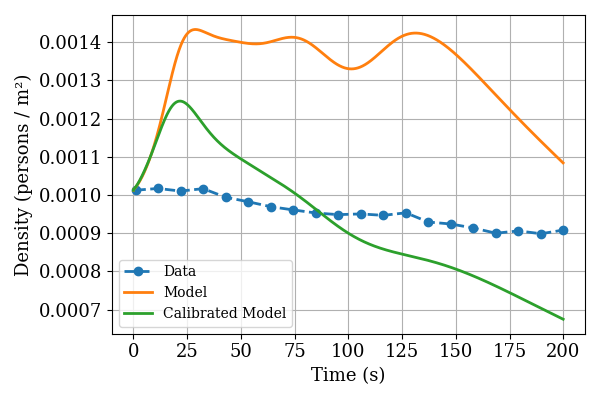}
	
	\caption{Time evolution of density at the spatial location $(-95.3, 67.5)$. Observed data are compared with model predictions obtained using the initial parameter set and the calibrated parameter set which is $\mathbf{\tilde{p}} = [7.23, 1.6, 13.13]$. The corresponding relative error (\ref{relative_error}) is $23.3\%$.}
	\label{total_density_-95.281_67.450}
\end{figure}

\begin{figure} [H]
	\centering
	\includegraphics[width=0.7\linewidth]{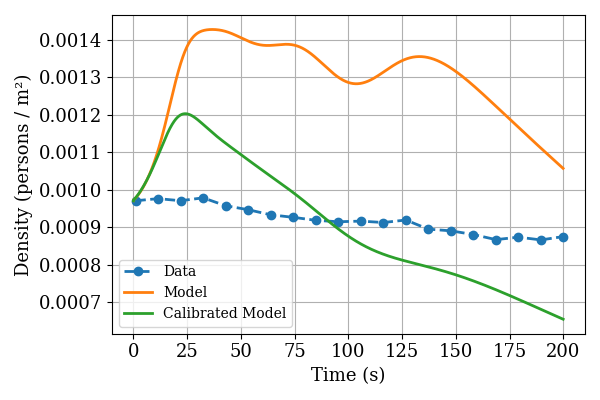}
	
	\caption{Time evolution of density at the spatial location $(-95.3, 68.9)$. Observed data are compared with model predictions obtained using the initial parameter set and the calibrated parameter set which is $\mathbf{\tilde{p}} = [7.23, 1.6, 13.13]$. The corresponding relative error (\ref{relative_error}) is $23.8\%$.}
	\label{total_density_-95.281_68.910}
\end{figure}

Figures \ref{total_density_-95.281_70.369}--\ref{total_density_-95.281_68.910} show the time evolution of the density, at three discrete spatial points, identified using the criterion described in Section \ref{Crowd_flow_calibration} via (\ref{choice_of_points}), where the observed data are compared with the model predictions obtained using the calibrated parameter set. %The time window corresponds to about one hour after the end of the match, and thus, the total densities obseved are in the range of $0.001 \text{ ped}/ \text{m}^2$. 
As in the previous section, the model provides a better aggreement with the observed smoothed data when the calibrated parameter set is used. The relative error corresponding to Figures  \ref{total_density_-95.281_70.369}--\ref{total_density_-95.281_68.910} is approximately $24\%$. 

\subsection{Calibration of the Complete Model}
To calibrate the epidemic spreading component of the model, we consider the time interval starting $50$ minutes after the end of the match, as it is described in Section \ref{crowd_flow_calibration_50_mins_after}, instead of the time window described in Section \ref{Crowd_flow_calibration}. The latter may not be suitable for estimating the parameters of the epidemic component of the model because, within that time interval, the number of exposed individuals remains constant, and thus the respective data may not be rich enough.  

We now proceed with the calibration of the complete coupled people-epidemic model, following the procedure and the setup described in Section \ref{Crowd_flow_calibration}. Since the epidemic component does not feed back into the crowd flow component (see Figure \ref{model_figure} and equations (\ref{contagion_model_1})--(\ref{beta_I})), the speeds and total density predicted by the crowd flow model remain invariant to changes in the epidemic spreading dynamics. Therefore, to calibrate the epidemic component, we use the speeds and total density predicted by the crowd flow component with the calibrated parameter vector $\mathbf{\tilde{p}}$. We employ the Nelder-Mead algorithm to estimate the epidemic parameter vector $\mathbf{\hat{p}} = [ i_0, r]$. The parameters are constrained to ensure that the calibrated values are physically meaningful, that is, small positive values for $i_0$ and positive values for $r$ (we note that, for example, in \cite{DiseaseContagion} $i_0=0.04$ and $r=1$), while yielding satisfactory and reasonable calibration results. The corresponding parameter intervals are 
\begin{equation}
	i_0 \in [0.0001, 0.1], \qquad r \in [0.01, 4].  
\end{equation}

\begin{figure} [H]
	\centering
	\includegraphics[width=\linewidth]{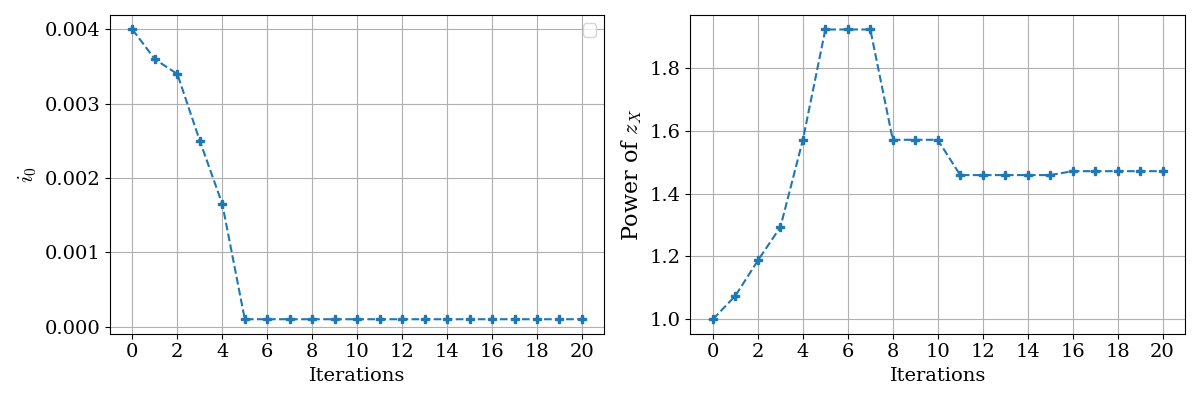}
	\caption{Convergence of the calibrated parameters $i_0$ and $r$, starting from the initial parameter set $\mathbf{\hat{p}}_{\text{init}}= [0.004, 1]$.}
	\label{i0_convergence}
\end{figure}

The cost function is modified as follows %\alpha, C_0, \tau,
\begin{equation} \label{cost_function_complete}
	J(\mathbf{\hat{p}})= \sqrt{\frac{1}{K N_x N_y} \sum_{k=0}^{K-1} \sum_{i=0}^{N_x-1} \sum_{j=0}^{N_y-1}  \left( \hat{b}_{\text{pred}}(i,j,k; \mathbf{\hat{p}}) - \hat{b}_{\text{obs}}(i,j,k) \right)^2 },
\end{equation}
 where $\hat{b}_{\text{pred}}(i,j,k; \mathbf{\hat{p}})=\rho^E(x_i,y_j,t_k; \mathbf{\hat{p}})$ denotes the model-predicted densities of exposed individuals using the parameter vector $\mathbf{\hat{p}}$ at the spatial location $(x_i, y_j)$ and at the discrete time $t_k$, with $t_k = \frac{10k}{ \Delta t}$ for $k=0, \dots, K-1$. The term $\hat{b}_{\text{obs}}(i,j,k) = \rho_{\text{obs}}^E(x_i,y_j,k)$ denotes the corresponding observed density of exposed individuals obtained from the smoothed data computed in Section \ref{ComputationOfMacroscopicDensitiesAndSpeeds}.  Although the densities of exposed, susceptible, and infected individuals are computed in Section \ref{Macroscopic_densities}, only the exposed density is included in the cost function (\ref{cost_function_complete}), because we observed (after exhaustive simulations) that this choice resulted in more satisfactory model calibration. This may be explained from the fact that including both quantities in the cost function may introduce conflicting optimization objectives, since the epidemic model predicts that every newly exposed individual corresponds exactly to subtraction of one susceptible individual. This relationship, however, may not hold exaclty in the numerical simulations due to numerical approximation errors introduced by the numerical scheme, as well as the sparse nature of the available data. The algorithm terminates when one of the following two criterion is satisfied. The first criterion, analogously to (\ref{1st_criterion}), is
\begin{equation} \label{1st_criterion_complete}
	\max_{i \in \{0,1,2\}} \Vert \mathbf{\hat{p}}_i - \mathbf{\hat{p}}^{\text{lowest}} \Vert_{\infty}< 0.1,
\end{equation}
where $\mathbf{\hat{p}}_i$ denotes the $i-th$ simplex vertex, and $\mathbf{\hat{p}}^{\text{lowest}}$ is the vertex corresponding to the lowest value of the cost function (\ref{cost_function_complete}). The second criterion is that the maximum number of 20 iterations is reached. As illustrated in Figure \ref{i0_convergence}, the parameter $i_0$ converges to its lower bound, while $r$ calibration does not activate the constraints. The former can be explained noting that the total number of exposed individuals predicted with the initial value of $i_0$ resulted in the model to significantly overestimate the total number of exposed individuals as compared with the data. 

\begin{figure} [H]
	\centering
	\includegraphics[width=0.9\linewidth]{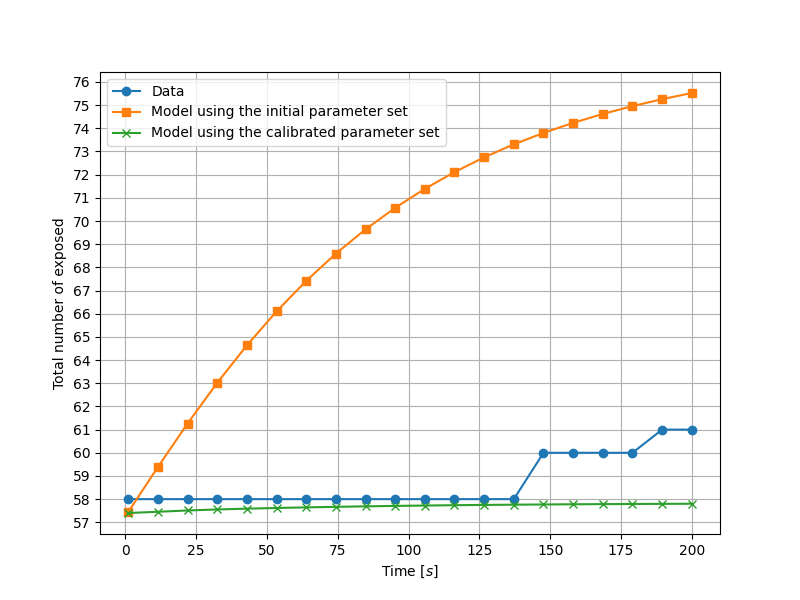}
	\caption{Comparison of the total number of exposed individuals obtained from the data, the model with the calibrated parameter set $\mathbf{\hat{\tilde{p}}}= [0.0001, 1.47]$, and the model using the initial parameter set $\mathbf{\hat{p}}_{\text{init}}= [0.004, 1]$.}
	\label{total_exposed}
\end{figure}

Figure \ref{total_exposed} shows the evolution of the total number of exposed individuals obtained from the data, the model using the calibrated parameter set \\ $\mathbf{\hat{\tilde{p}}}= [0.0001, 1.47]$, and the model using the initial parameter set $\mathbf{\hat{p}}_{\text{init}}= [0.004, 1]$. As expected, the calibrated model provides a better agreement with the data, with a maximum (over time) relative error of $5.3\%$, while the non-calibrated model resulted in a relative error of $26.4\%$. We note that due to the sparsity of the available data and the effect of the necessary smoothing applied, the total number of exposed individuals provides a more consistent, overall illustration of the calibration results, as compared with the local variations of exposed density at specific (microscopic) spatial locations. 

\section{Conclusions and Future Work} \label{Remarks}
In this work, we derived trajectories of exposed-susceptible-infected individuals from microscopic data and we computed the corresponding macroscopic densities and velocities, together with their smoothed counterparts. We then presented and solved numerically a coupled, PDE crowd flow - epidemic spreading model using a finite volume scheme. Utilizing the computed macroscopic quantities and the numerical scheme, we performed a calibration procedure for obtaining model parameters that provide an optimal matching between the densities obtained from the data and from the model. The calibration results indicated that the proposed procedure, on average, resulted in 24\% relative error, which is deemed sufficient in view of the sparsity of the data used.

Due to the sparsity of the data, although pedestrian speeds were computed, they were not included in the calibration procedure, because, despite the smoothing applied, the obtained speeds remained insufficiently representative of actual speeds, thus essentially introducing additional uncertainty in the calibration procedure. The sparsity of the data used and the unavailability in general of detailed density/speed data over a spatiotemporal scale of meters-hours in conjuction with the corresponding epidemic spreading data, calls for development of a calibration procedure employing fictitious epidemiological/trajectory data generated by microscopic simulators as the ones in, for example, \cite{Activity_based_epidemic_propagation, LBS_agent_based_simulator}. This can be pursued as future research.

\section*{Acknowledgments}
Funded by the European Union (ERC, C-NORA, 101088147). Views and opinions expressed are however those of the authors only and do not necessarily reflect those of the European Union or the European Research Council Executive Agency. Neither the European Union nor the granting authority can be held responsible for them.
%\appendix
%\section{}

%%       \cite{<label>} ==> [1]

%%
%Example citation, See \cite{lamport94}.

 \bibliographystyle{elsarticle-num} 
 \bibliography{elsarticle-template-num}




%% For numbered reference style

%%\bibitem{lamport94}
%%  Leslie Lamport,
%%  \textit{\LaTeX: a document preparation system},
%% Addison Wesley, Massachusetts,
%%  2nd edition,
%%  1994.

%%\end{thebibliography}
\end{document}